\documentclass[pre,preprint,showpacs]{revtex4}
\usepackage{epsfig}
\usepackage{amssymb}
\usepackage{amsfonts}
\usepackage{amsthm}
\usepackage{newlfont}
\usepackage{epsfig}
\usepackage{amscd}
\usepackage{graphics}
\newcommand{\NN}{{\mathbb N}}

\newcommand{\beq}{\begin{equation}}
\newcommand{\eeq}{\end{equation}}
\newcommand{\ba}{\begin{array}}
\newcommand{\ea}{\end{array}}
\newcommand{\bea}{\begin{eqnarray}}
\newcommand{\eea}{\end{eqnarray}}
\begin{document}
%\begin{frontmatter}

\title{Relationship between probabilities of the state transfers and entanglements in spin systems with simple geometrical configurations.}
\author{S.I.Doronin, E.B.Fel'dman and A.I.Zenchuk}
\email{ efeldman@icp.ac.ru, zenchuk@itp.ac.ru}
\affiliation{
Institute of Problems of Chemical Physics, Russian Academy of Sciences, Chernogolovka, Moscow reg., 142432, Russia}

\date{\today}

\begin{abstract}
In this paper we derive analytical  relations between  probabilities  of  the excited state transfers  and  entanglements calculated by both the Wootters and positive partial transpose (PPT) criteria   for the arbitrary spin system  with single excited spin in the external magnetic field and Hamiltonian commuting with $I_z$. We apply these relations to study the 
arbitrary state transfers and entanglements  in the
simple systems of nuclear spins having two- and three-dimensional geometrical configurations with $XXZ$ Hamiltonian. It is shown that High-Probability State Transfers (HPSTs)  are possible among all four nodes placed in the corners of the rectangle with the proper ratio of sides as well as among all eight nodes placed in the corners of the parallelepiped with the proper ratio of sides. Entanglements responsible for these HPSTs have been identified. 
\end{abstract}


\maketitle

\section{Introduction}

This paper is devoted to the problem of the high probability state transfer (HPST) \cite{KZ} among many nodes of the spin system and to the  relationship between probabilities of HPSTs   and entanglements responsible for these 
transfers. We consider  
  nuclear spin-1/2 systems with XXZ-Hamiltonian and  different geometrical configurations. However, the above relationship remains valid for any Hamiltonian ${\cal{H}}$  commuting with the total projection operator $I_z$.

By "state transfer" we mean the following phenomenon \cite{Bose,CDEL}.
Consider the   chain of spin-1/2 with dipole-dipole interaction 
in the strong external magnetic field.  Let  all spins be 
directed along the external magnetic  field except the $i$th one 
whose initial state is arbitrary. In other words, let the spin 
system be prepared in the state  $\psi_{ii}= \cos (\theta/2) 
|0\rangle + e^{-i\phi} \sin (\theta/2) |i\rangle$, where 
$|0\rangle$ is a ground state, i.e. all spins are directed along 
the magnetic field and $|i\rangle $ means that only $i$th spin is
directed opposite to the external magnetic field (i.e. $i$th spin
is excited).  Let the energy of the ground state be zero. If  the
state becomes  $\psi_{ij}= \cos (\theta/2) |0\rangle + e^{-i\phi}
\bar f_{ij} \sin (\theta/2) |j\rangle$ with $|\bar f_{ij}| = 1$ 
at the time moment  $t=t_{ij}$ then we say that the initial state
has been  transferred from the $i$th to the $j$th node with the 
phase shift $\bar \Gamma_{ij}=\arg \bar f_{ij}$. Since $|\bar 
f_{ij}|=1$, all other spins are directed along the field at 
$t=t_{ij}$.  Here $\bar f_{ij}=f_{ij}(t_{ij})$ and $f_{ij}(t)$ is
the transition amplitude  of  an excited state 
from the $i$th to the $j$th node: $
f_{ij}(t)=\langle j |e^{-i{\cal{H}} t}|i\rangle$.  
% Note, that the phase shifts $\Gamma_{ij}$ do not depend on both %parameters $\phi$ and $\theta$ of the initial state and may be 
%found for each particular chain. 
It is known \cite{Bose} that the  effectiveness of the state transfer between the  $i$th and $j$th nodes  may be characterised by the fidelity $F_{ij}(t)$ 
 \begin{eqnarray}\label{F_nm}\label{fidelity}
F_{ij}(t)=\frac{|f_{ij}(t)| \cos \Gamma_{ij}(t) }{3} + \frac{|f_{ij}(t)|^2}{6} +\frac{1}{2}.
\end{eqnarray}
%where the amplitudes $f_{ij}$ and the phases $\Gamma_{ij}$ are %defined as follows 
%\begin{eqnarray}\label{f_nm}
%\Gamma_{ij}(t)={\mbox{arg}} f_{ij}(t)
%.
%\end{eqnarray}
We see that the fidelity is maximal for $\Gamma_{ij}=0$ ${\mbox{mod}} \;2\pi$.
 If the external magnetic field is homogeneous and we are interested in the state propagation between two nodes, say between  $s$th and $r$th nodes at the moment $t_{rs}$, then  condition $\bar \Gamma_{rs}\equiv \Gamma_{rs}(t_{rs})=0$  may be simply satisfied by the proper choice of the constant magnetic field value  \cite{Bose}. In this  case the fidelity $\bar F_{rs}\equiv F_{rs}(t_{rs})$ takes maximal value together with absolute value of the  transition amplitude $|\bar f_{rs}|\equiv |f_{rs}(t_{rs})|$. For this reason, namely  $|\bar f_{rs}|$  (rather then $\bar F_{rs}$)  is considered as the characteristic of the state transfer in many refs, see,  
 for instance, \cite{GMT,TPVH,CDEL}. It is clear that  
 $|f_{rs}(t)|$ may be replaced by the probability of the excited state transfer $P_{rs}(t)=|f_{rs}(t)|^2$ and 
 $\bar P_{rs}=|\bar f_{rs}|^2$ \cite{KZ,FZ}.
 
 More general case of the HPSTs  among many nodes of the $N$ node spin chain  has been studied in \cite{FZ}. In this case any particular state transfer between the $i$th and $j$th nodes is associated with its own phase shift $\bar \Gamma_{ij}$, $i,j=1,\dots,N$ (note that $i$ may be equal to $j$ which means return of the state to the $i$th node).  However, it is important that all these shifts may be eliminated using magnetic field properly depending on time \cite{FZ}. 
For this reason 
the effectiveness of the state transfer between the $i$th and the $j$th nodes may be equivalently described either by the  fidelity $\bar F_{ij}$  or by the probability  of the excited state transfer $\bar P_{ij}$ even in this generalized case. Namely optimization of $\bar P_{ij}$ allows us to find all necessary parameters of the geometrical spin configuration providing HPSTs among many nodes while  phases  $\bar \Gamma_{ij}$ may be removed by the appropriate time dependent external magnetic field as it was done in \cite{FZ}. For this reason,  we will study the probability of the  single excited state transfer instead of the fidelity (\ref{fidelity}) of arbitrary state transfer in the subsequent sections of this paper. This means that the spin 1/2 system of $N$ nodes  is prepared in the initial state  
\begin{eqnarray}\label{initial}
\psi_{ii}=|i\rangle,
\end{eqnarray}
 where $i$ takes one of the values  $i=1,\dots,N$.

  The problem of the perfect state transfers (PSTs), HPSTs and 
  entanglements in the spin systems is  very attractive and 
different aspects of this problem have been studied in many 
details \cite{FBE,CDEL,GKMT,GMT,ACDE,FR,KF,VGIZ,KS,BGB,BCMS,TPVH}. Nevertheless, most of the results  
are devoted to 
the  linear and circular spin  chains, which are considered as communication 
channels in the quantum information systems.  Different 
Hamiltonians describing these chains have been studied, such as 
$XY$, $XYZ$, Heisenderg Hamiltonians. Usually, the approximation of 
the nearest neighbour interactions is taken as 
a basic tool for such studies. Note, that this is a good 
approximation in the case of, for instance,  exchange 
interaction, when coupling constants decrease exponentially with increase in 
the distance.  However, this approximation is not satisfactory 
for the spin systems with dipole-dipole interaction (such as 
nuclear spin systems in solids) and a wide spread of the coupling constants.  
  
Most efforts have been turned to the study of two  phenomena: the state transfer along  chains \cite{Bose,CDEL,GMT,TPVH,FBE,GKMT,KF,KS} and two-qubit entanglements in  chains (such as entanglement between end nodes or between nearest nodes \cite{Bose,GKMT,VGIZ,DPF}). The Wootters criterion is applicable in this case  allowing one to describe the  entanglement in terms of so-called concurrence \cite{HW}.  It is important that there is an analytical  dependence of the concurrence  between two nodes on the probability of the state transfer between these nodes  which was derived  in \cite{AOPFP,GMT} for the system with single excited spin. However, complicated system of $N$ spins  exhibits entanglements not only between two nodes, but also between arbitrary  two subsystems. These entanglements may be effectively described  by the positive partial transpose  (PPT) criterion \cite{P,VW} introducing so-called double negativity as a measure of entanglement. Explicit relations of the double negativity associated with two subsystems on  the probabilities of the excited state transfers between different nodes  will be derived in this paper for the spin system with single excitation.

%Usually, the excited state prepared in the $i$th  node ($t=0$) %is considered to be transfered to the $j$th node at the moment %$t=t_0$ if the probability to register  the excited state in %$j$th node at the moment $t_0$  is one (PST) %\cite{CDEL,ACDE,KS}. 
Although the PST would be preferable in the quantum communication chains, 
it is hardly realizable in experiments with  long chains  because of the following two basic reasons:
\begin{enumerate}
\item
Theoretical prediction of the PST in the long chains  is associated with the approximation of the  Hamiltonian by the  nearest neighbour interaction, while the complete Hamiltonian  must be used in practice. As we have already noted, this approximation is well applicable to the systems with exchange interaction and is not always valid for the systems with  dipole-dipole interaction. 
\item
Coupling constants may not be always known as accurately as we want in the case of both exchange and dipole-dipole interaction.
\end{enumerate}
Thus, HPST between different nodes of the spin system \cite{KZ} seems to be more realistic in comparison with PST. 

In this paper we will study the spin systems with dipole-dipole interaction  in the external magnetic field with single excited node described by the $XXZ$ Hamiltonian. We will study  the spin systems with different 
 geometrical configurations of  nodes and  initial state (\ref{initial}) 
which may  provide  the HPSTs of the excited state among all of nodes. We will show which parts of the spin system must be entangled in order to provide each of these transfers.
 The study of two- and three-dimensional spin systems is important because they  are more compact and consequently  they are   more promising  as quantum registers and/or short communication channels. It will be shown in this paper, that namely such configurations (more precisely, spin configurations with nodes placed in the corners of either rectangle or parallelepiped) provide   the HPSTs among several different nodes of the spin system  during relatively short time interval in comparison with the line systems \cite{FZ} which is important  for the development of the quantum information systems and/or short communication channels.  Our study is also stimulated by the  experiments 
 %, for instance,  the multiple quantum coherence intensities 
 on the   quantum information processes  
 in  solids 
 %cubic crystals such as $CaF_2$.
%and $FAp$ 
%\cite{CEBRLC}.
\cite{K-S,CCCR}. 

 This paper is organized as follows.
In Sec.\ref{Section:XYZ} (and in Appendices \ref{App:Wootters} and \ref{App:PPT}) we obtain analytical  dependence of  the either  concurrences (Wootters criterion) or  double negativities (PPT criterion)  between  different two subsystems of the spin system    on the probabilities of the state transfers, generalizing the results of refs.\cite{AOPFP,GMT}. 
In Sec.\ref{Section:two_nodes}  we consider the simplest one-dimensional model of two nodes where the relationship between  entanglement and probability of the state transfer is most transparent and an equivalent result may be obtained using either the Wootters \cite{GMT} or PPT criterion.  Two-dimensional spin systems  will be considered in Sec.\ref{Section:four_nodes}, see also Appendix \ref{App:4nodes}.   We  arrange HPST among all nodes of the four-node spin system (rectangular geometry) and show that the external magnetic field directed along one of the sides of the rectangle decreases significantly (more then twice) the time intervals needed for the HPSTs among nodes in comparison with the case when the field is perpendicular to the plane of the rectangle. 
 Similar study of the three-dimensional eight-node system (with spins placed in the corners of the parallelepiped) is represented in  Sec.\ref{Section:eight_nodes}, see also Appendix \ref{App:8nodes}.
It is evident that HPSTs may not be effectively arranged in the arbitrary system of nodes. Detailed  algorithm allowing one to obtain  parameters of the rectangle spin system (namely, the ratio of sides of the rectangle) with the  HPSTs among all  four  nodes  is given in Appendix \ref{App:4nodes}. Particular example of the eight-node three-dimensional spin system with HPSTs among all nodes (parallelepiped configuration) is represented in Appendix \ref{App:8nodes}. 

%%%%%%%%%%%%%%%%
\section{Spin-1/2 systems with single excited node   and $XXZ$ Hamiltonian }
\label{Section:XYZ}
We study the HPSTs and entanglements   among  the nodes of the  spin-1/2 system in the  external magnetic field described by the $XXZ$
 Hamiltonian with zero Larmor frequencies:
\begin{eqnarray}\label{Hamiltonian}
&&
{\cal{H}}=\sum_{{i,j=1}\atop{j>i}}^{N}
D_{ij}(I_{i,x}I_{j,x} + I_{i,y}I_{j,y}-2  I_{i,z}I_{j,z}),\\\label{Dij}
&&
D_{ij}=\frac{1-3 \cos^2 \theta_{ij} }{r_{ij}^3}\gamma^2 \hbar,
\end{eqnarray}
where $\gamma$ is gyromagnetic ratio,  $r_{ij}$ is the distance between $i$th and $j$th spins, $\theta_{ij}$ is the angle between the external magnetic field and $r_{ij}$, $I_{i,\alpha}$ is the projection operator of the $i$th  spin on the $\alpha$ axis, $\alpha=x,y,z$,
$D_{ij}$  are the dipole-dipole coupling constants.
This Hamiltonian describes  the secular part of the dipole-dipole interaction in the strong external  magnetic field
\cite{A}.
We denote $D_n\equiv D_{n,n+1}$, $n=1,\dots,N-1$. Taking into account the definition of $D_{ij}$,
for description of the spin system with arbitrary geometrical configuration we use the coordinates of each node, i.e. the set of the following triads
\begin{eqnarray}
(x_i,y_i,z_i),\;\;i=1,\dots,N,
\end{eqnarray}
so that 
\begin{eqnarray}
&&
r_{ij}=\sqrt{(x_j-x_i)^2 + (y_j-y_i)^2 + (z_j-z_i)^2},\\\nonumber
&&
\cos^2\theta_{ij}= \frac{(z_j-z_i)^2}{r_{ij}^2}.
\end{eqnarray}
It is important, that the Hamiltonian (\ref{Hamiltonian}) commutes with $I_z$ ($z$-projection of the total spin):
\begin{eqnarray}
[{\cal{H}},I_z]=0.
\end{eqnarray}
This means that both ${\cal{H}}$ and $I_z$ have the common set of eigenvectors. It is convenient to write the eigenvectors of the operator $I_z$ in terms of the Dirac notations.  Let
\begin{eqnarray}\label{basis}
|n_1\dots n_N \rangle
\end{eqnarray}
be the eigenvector of the operator $I_z$ where  the $i$th spin is directed opposite to the external magnetic field if $n_i=1$ and  along the field if $n_i=0$. For the sake of brevity, hereafter we will use notations $|0\rangle$ for the eigenvector associated with the state when all spins  are directed along the external field and   $|i_1\dots i_k\rangle$ for the eigenvectors associated with the state when  $i_1$th, $\dots$, $i_k$th spins are directed opposite to the external field, i.e. these spins are excited. Thus eigenvector $|i\rangle$ means that only $i$th spin is excited.  
Using these notations, the basis (\ref{basis}) may be ordered as follows:
\begin{eqnarray}\label{basis_ordered}
&&
|i_1\rangle,  \;\;|0\rangle, \;\;|i_1i_2\rangle, \;\;\dots,\;\;|i_1\dots i_N\rangle,  \;\;i_1<i_2<\dots<i_N,\;\;{\mbox{i.e.}}\\\nonumber
&&
 i_1=1,\dots,N,\;\;i_k=i_{k-1}+1,\dots,N,\;\;k=2,\dots,N.
\end{eqnarray}

The matrix representation $H$ of the Hamiltonian ${\cal{H}}$  in  basis (\ref{basis_ordered}) gets the following diagonal block structure:
\begin{eqnarray}\label{H_diag}
{{H}}={\mbox{diag}}(H_1,H_0,H_2,H_3\dots,H_N),
\end{eqnarray}
where the block $H_i$ is associated with the set of states of the whole spin system 
 having $i$ spins directed opposite to the field.

Hereafter we will study the problem of the  single excited quantum state transfer among nodes of  the spin-1/2  system  with the XXZ Hamiltonian in the external magnetic field. We say, that  the $k_0$th node is excited initially. 
It is important that only the block $H_1$  is nonzero in this case:
%, we may reduce the $2^N$-dimensional basis (\ref{basis}) to the %following $N$-dimensional one:
% \begin{eqnarray}\label{basis1}
%|i\rangle,\;\;i=1,\dots,N,
%\end{eqnarray} 
%which means that 
%only the block $H_1$ remains in the matrix representation of the %Hamiltonian,
% i.e.
  \begin{eqnarray}\label{H1}
&&
H_1=\frac{1}{2}(D -\Gamma I),\\\nonumber
&&
D=\left(\begin{array}{ccccccc}
A_{11} & D_1 &D_{13} & \cdots& D_{1(N-2)} & D_{1(N-1)}&D_{1N}\cr
D_1& A_{22} & D_2 & \cdots& D_{2(N-2)} & D_{2(N-1)}&D_{2N}\cr
D_{13}&D_2& A_{33} & \cdots& D_{3(N-3)} & D_{3(N-1)}&D_{3N}\cr
\vdots &\vdots &\vdots &\vdots&\vdots &\vdots &\vdots \cr
D_{1(N-2)}&D_{2(N-2)}&D_{3(N-2)}&\cdots&A_{(N-2)(N-2)}&D_{j-1}&D_{(N-2)N}\cr
D_{1(N-1)}&D_{2(N-1)}&D_{3(N-1)}&\cdots&D_{j-1}&A_{(N-1)(N-1)}&D_j\cr
D_{1N}&D_{2N}&D_{3N}&\cdots&D_{(N-2)N}&D_j&A_{NN}
\end{array}\right),
\\\nonumber
&&
A_{nn}=2\sum_{{i=1}\atop{i\neq n}}^{N}D_{in},\;\;\Gamma=\sum_{{i,j=1}\atop{i< j}}^{N}D_{ij},
\end{eqnarray}
where $I$ is $N\times N$ identity matrix.
This simplification of the Hamiltonian allows one to
\begin{enumerate}
\item
derive explicit analytical dependence of concurrence and/or double negativity (as  measures of entanglement between any two subsystems of the spin system) on the probabilities  of the state transfers between different nodes of the system, which is hardly realizable in the case of the Hamiltonian with general structure (\ref{H_diag});
\item
perform the numerical simulations of the state transfers in the big spin systems, which is hardly realizable  in general case \cite{DFGM}.
\end{enumerate} 
Hereafter we will use the dimensionless  time $\tau$,  
 coupling constants $d_{nm}$ and distances $\xi_{nm}$,
\begin{eqnarray}\label{tau}
&&
\tau= D_{12} t,
\;\;\;
d_{nm}=\frac{D_{nm}}{D_{12}},\;\;\;
\xi_{nm}=\frac{r_{nm}}{r_{12}}.
\end{eqnarray}
Using  definitions (\ref{tau}) and taking into account that $D_{12}=\gamma^2 \hbar/r_{12}^3$ in all our examples,  Hamiltonian (\ref{Hamiltonian}) may be written as follows:
\begin{eqnarray}\label{Hamiltonian_dl}
&&
{\cal{H}}=D_{12} \tilde {\cal{H}},\;\;\;
\tilde {\cal{H}}=\sum_{{i,j=1}\atop{j>i}}^{N}
d_{ij}(I_{i,x}I_{j,x} + I_{i,y}I_{j,y}-2  I_{i,z}I_{j,z}),\\\label{Dij_dl}
&&
d_{ij}=\frac{1-3 \cos^2 \theta_{ij} }{\xi_{ij}^3}.
\end{eqnarray}

%%%%%%%%%%%%%%%%
\section{Probabilities of the state transfers and entanglements between different subsystems of the arbitrary nuclear spin system}

First of all, in order to establish the relationships between probabilities of the state transfers among nodes  and   entanglements among them   we will need density matrix $\rho$ introduced as follows:
\begin{eqnarray}
\rho=e^{-i\tilde {\cal{H}} \tau} |k_0\rangle \langle k_0 | e^{i\tilde {\cal{H}} \tau},
\end{eqnarray}
where  $k_0$ means that $k_0$th spin  was directed opposite to the field initially. Only such initial states will be considered hereafter.
It is important,  that this matrix may be  written in the following block form using  basis (\ref{basis_ordered}) 
\begin{eqnarray}
\rho=\left(
\begin{array}{cc}
A&0_{N,2^N-N}\cr
0_{2^N-N,N} & 0_{2^N-N,2^N-N}
\end{array}
\right),\;\;\;A=\left(\begin{array}{ccc}
a_{11}&\cdots & a_{1N}\cr
\cdots&\cdots&\cdots\cr
a_{N1}&\cdots & a_{NN}
\end{array}
\right),
\end{eqnarray}
where $0_{n,m}$ means $n\times m$ zero matrix
 and nonzero elements  $a_{ij}$ are defined as follows:
 \begin{eqnarray}
 a_{ij}\equiv \langle i|\rho|j\rangle = f_{k_0i} f^*_{k_0j},\;\;a_{ji}=a^*_{ij},\;\;i,j=1,\dots,N,
 \end{eqnarray}
where $f_{nm}$ are transmission amplitudes,
 \begin{eqnarray}
 f_{nm}&=&\langle m |e^{-i\tilde {\cal{H}} \tau}|n\rangle=\sum_{j=1}^{N}  u_{nj} u_{mj}e^{-i\lambda_j  \tau/2},\;\;f_{nm}=f_{mn}.
 \end{eqnarray}
 Here $u_{ij}$, $i,j,=1,\dots,N$, are the components of the normalized eigenvector
$u_j$ corresponding to the  eigenvalue $\lambda_j$ of the matrix
$D$:
$Du_j=\lambda_j u_j$.

%%%%%%%%%%%%%
\subsection{Probability of the state transfer from the $k_0$th to the $m$th node of the $N$-node spin chain}
The probability  $P_{k_0m}$ of the state transfer from the $k_0$th to the $m$th node as  a function of time  is defined by the diagonal element $a_{mm}$ of the density matrix. In fact 
\cite{KF},
\begin{eqnarray}\label{P^N}
P_{k_0m}(\tau)&=&\left|\langle m |e^{-i{\tilde {\cal{H}}} \tau}|k_0\rangle
\right|^2=\left|f_{k_0m}
\right|^2\equiv a_{mm}.
\end{eqnarray}

Throughout this paper we will use notations  $\bar P_{k_0m}$ and  $\tau_{k_0m}$ for the  probability of the HPST  and  for the time interval required for the  HPST  from the $k_0$th to the $m$th node of the $N$-node chain:
\begin{eqnarray}
\bar P_{k_0m}\equiv P_{k_0m}( \tau_{k_0m}).
\end{eqnarray}
By our definition, the state transfer from the $k_0$th node to  the $m$th node  will be referred to as HPST if
\begin{eqnarray}\label{HPST}
\bar P_{k_0m}\ge P_0.
\end{eqnarray}
The value  $P_0$ is conventional, in our paper we take $P_0=0.9$.
In addition, there is an  important parameter of the HPSTs, namely the time interval ${\cal{T}}$ during which the excited state    may be detected in all nodes of the system \cite{KZ}: 
\begin{eqnarray}\label{tT}
{\cal{T}}=\max_{n=1,\dots,N} \tau_{k_0n}.
\end{eqnarray}

%%%%%%%%%%%
\subsection{Wootters criterion: two-node entanglements in the spin system}
It is well known that the entanglement between two nodes $i$ and $j$  of the $N$-node spin system  may be described by the Wootters criterion \cite{HW}, which introduces the so-called concurrence $C_{ij}$ as a measure of the entanglement:
\begin{eqnarray}\label{C_ij}
C_{ij}=\max\left(0,2\lambda-\sum_{n=1}^{4}\lambda_n\right),\;\;
\lambda=\max(\lambda_1,\lambda_2,\lambda_3,\lambda_4).
\end{eqnarray}
Here $\lambda_n$, $n=1,2,3,4$ are the square roots of the eigenvalues of the $4\times 4$ matrix $\hat\rho_{ij}$
\begin{eqnarray}\label{hat_rho}
 \hat\rho_{ij}= (\tilde\rho^{red}_{ij})^*\rho^{red}_{ij},
 \end{eqnarray}
where $*$ means complex conjugate, $\rho^{red}_{ij}$ is the reduced density matrix, i.e.
\begin{eqnarray}\label{rho_red_ij}
\rho^{red}_{ij} = {\mbox{Tr}}_{{n=1,\dots,N}\atop{n\neq i,j}} \rho.
\end{eqnarray}
 Matrix $\tilde \rho^{red}_{ij}$ is defined as
  \begin{eqnarray}
  \tilde \rho^{red}_{ij}=\sigma^y_i\otimes\sigma^y_j \rho^{red}_{ij}\sigma^y_j \otimes\sigma^y_i.
  \end{eqnarray}
After simple calculations (see Appendix \ref{App:Wootters} for details) one derives the following formula:
\begin{eqnarray}\label{CP}
C_{ij}=2 |a_{ij}|=2\sqrt{P_{k_0i} P_{k_0j}},\;\;i\neq j. 
\end{eqnarray}
This relation is valid for the system with any number of spins and for any Hamiltonian commuting with $I_z$ \cite{AOPFP,GMT}.

%%%%%%%%%%%%%%%
\subsection{PPT criterion: entanglement between two arbitrary subsystems}
PPT criterion describes the entanglement between any two subsystems $A$ and $B$ of the system $S$ in terms of the so-called double negativity $N_{A,B}$ \cite{VW,ZHSL}, which is the absolute value of the doubled sum of the negative eigenvalues of the matrix $\rho_{AB;C}$,
\begin{eqnarray}\label{PPT}
\rho_{AB;C}=(\rho_{A,B})^{T_A},\;\;\;S=A\cup B \cup C,
\end{eqnarray}
where the reduced density matrix $\rho_{A,B}$ is defined as follows:
\begin{eqnarray}\label{DN_rho_red}
\rho_{A,B}=Tr_{C}\rho,
\end{eqnarray}
and $T_A$ means the transposition with respect to the subsystem $A$. In 
particular, the subsystem $C$ may be empty.
It is important that one can write the explicit formulae for 
$N_{A,B}$  for the spin system in the magnetic field  with single excited spin described by any Hamiltonian commuting with the total projection $I_z$ (see Appendix \ref{App:PPT} for details):
\begin{eqnarray}\label{PPT1}
N_{i_1\dots i_{M_1},j_1\dots j_{M_2}}&=&- \left( \sigma_{i_1\dots i_{M_1}j_1\dots j_{M_2}}-\sqrt{\sigma_{i_1\dots i_{M_1}j_1\dots j_{M_2}}^2+4\sum_{n=1}^{M_1}\sum_{m=1}^{M_2} |a_{i_nj_m}|^2}\right)=\\\nonumber
&&
- \left( \sigma_{i_1\dots i_{M_1}j_1\dots j_{M_2}}-\sqrt{\sigma_{i_1\dots i_{M_1}j_1\dots j_{M_2}}^2+4\sum_{n=1}^{M_1} P_{k_0i_n} \sum_{m=1}^{M_2} P_{k_0j_m}}\right),
\end{eqnarray}
where $A=\{i_1\dots i_{M_1}\}$, $B=\{j_1\dots j_{M_2}\}$,
\begin{eqnarray}
\sigma_{i_1\dots i_{N_0}}=\sum_{{n=1}\atop{n\neq i_1,\dots,i_{N_0}}}^N a_{nn}=\sum_{{n=1}\atop{n\neq i_1,\dots,i_{N_0}}}^N P_{k_0n}=1-\sum_{n=1}^{N_0} P_{k_0n}
\end{eqnarray}
(we use the identity $\sum_{n=1}^N P_{k_0n}\equiv 1$).
In particular,
\begin{eqnarray}
\label{PPT2}
N_{i,j}&=&-\left( \sigma_{ij}-\sqrt{\sigma_{ij}^2+4 |a_{ij}|^2 }\right)=
-\left( \sigma_{ij}-\sqrt{\sigma_{ij}^2+4 P_{k_0i} P_{k_0j}}\right)=\\\nonumber
&&
-\left(1-P_{k_0i}-P_{k_0j}-
\sqrt{(1-P_{k_0i}-P_{k_0j})^2+4 P_{k_0i} P_{k_0j}}\right),\\\label{PPT3}
N_{i,rest}&=&2\sqrt{\sum_{{j=1}\atop{j\neq i}}^N |a_{ij}|^2}=
2\sqrt{P_{k_0i}\sum_{{j=1}\atop{j\neq i}}^N P_{k_0j}}=
2\sqrt{P_{k_0i}(1-P_{k_0i})},\\\label{PPT4}
N_{i_1i_2,j_1j_2}&=&-\left(
\sigma_{i_1i_2j_1j_2}-\sqrt{\sigma_{i_1i_2j_1j_2}^2+4\sum_{n=1}^2 \sum_{m=1}^2 |a_{i_nj_m}|^2}
\right)=\\\nonumber
&&-\left(
\sigma_{i_1i_2j_1j_2}-\sqrt{\sigma_{i_1i_2j_1j_2}^2+4(P_{k_0i_1}+P_{k_0i_2})(
P_{k_0j_1}+P_{k_0j_2})}\right).
\end{eqnarray}
where $N_{i,rest}$ is double negativity associated with the entanglement between $i$th spin and the rest part of the spin system.
It has the most simple expression  depending only on $P_{k_0i}$. 
%For this reason, the value  $N_{i,rest}$ may not be considered 
%as a meaningfull  characteristics of the entanglement for this 
%type of processes.
%It takes the maximal value at $P_{k_0i}=0.5$: %$(N)_{i,rest})_{max}=1$. 

%The  relations (\ref{PPT1}-\ref{PPT4}) justify the 
%following inequalities \cite{VW}:
%\begin{eqnarray}
%N_{i_1,j_1}\le N_{i_1,j_1j_2} \le\dots\le  N_{i_1,j_1\dots %j_{M}}.
%\end{eqnarray} 

 Eqs.(\ref{PPT1},\ref{PPT2}-\ref{PPT4}) are simple relations between two important characteristics of the spin system: entanglement between arbitrary subsystems $A$ and $B$ (characterized by the double negativity) and  probabilities of the state transfers from the $k_0$th node to other nodes of the subsystems $A$ and $B$. However, since $(a)$  both these characteristics have rather complicated oscillating behaviour and $(b)$ it is difficult to "inverse" relation  (\ref{PPT1}) (i.e. to write probabilities as functions of double negativities), it is hard to get  answers to the following questions: 
 \begin{enumerate}
\item
 Which geometrical configurations of the spin system provide HPSTs among all nodes?
 \item
 Which entanglements are responsible for the HPSTs among  different nodes of the spin system?
 \end{enumerate}
We will answer these questions for the particular systems of two, four and eight nodes in Secs.\ref{Section:two_nodes}-\ref{Section:eight_nodes}.

%%%%%%%%%%%%%%
\section{Perfect state transfer and entanglement in the system of two nodes}
\label{Section:two_nodes}

\begin{figure*}[!htb]
\noindent    
\resizebox{70mm}{!}{\includegraphics[width=5cm,
angle=270]{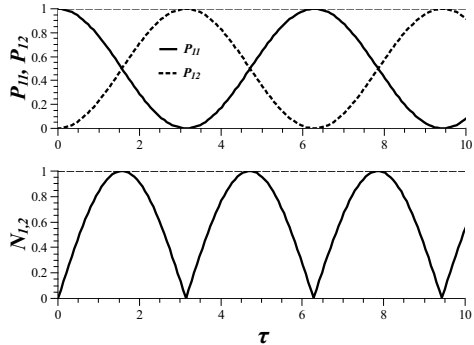}
%{2qubit.eps}
}
  \caption{The probabilities  $P_{11}$ and $P_{12}$ and the double negativity $N_{1,2}$ in the two-node spin system
  }
  \label{Fig:2qubit}
\end{figure*}

It was shown in \cite{CDEL} that the probabilities for the excited state to be detected  in the first and in the second nodes are following ($k_0=1$):
\begin{eqnarray}
&&
P_{11}(\tau)=\cos^2(\tau/2),\;\;P_{12}(\tau)=\sin^2(\tau/2),
\end{eqnarray}
while the double negativity $N_{1,2}$ is defined by eq.(\ref{PPT2}):
\begin{eqnarray}
N_{1,2}(\tau)\equiv C_{12}(\tau)=2\sqrt{P_{11}(\tau)P_{12}(\tau)}=|\sin \tau|,
\end{eqnarray}
i.e. $\bar P_{12}=1$ at $\tau_{12}=\pi +2\pi n$, $n=0,\pm 1,\pm 2,\dots$.
We compare the graphs of $P_{11}(\tau)$ and $P_{12}(\tau)$ with double negativity $N_{1,2}(\tau)$, see Fig.\ref{Fig:2qubit}, and observe that maxima of $N_{1,2}$ correspond to $P_{11}=P_{12}=1/2$ while the minima of $N_{1,2}$ correspond to the maxima  and  minima of  $P_{1i}$, $i=1,2$.
This result is not surprising. In fact, 
the wave function associated with the two-node spin system is following:
\begin{eqnarray}
\Psi(\tau)=f_{11}(\tau) |10\rangle +  f_{12}(\tau) |01\rangle,\;\;\;
P_{11}=|f_{11}|^2,\;\;P_{12}=|f_{12}|^2. 
\end{eqnarray} 
However, if $P_{11}=0$ or $P_{12}=0$ this function reduces to 
$\Psi= |01\rangle$ or $\Psi= |10\rangle$ respectively. 
It is well known that these states are separable and their  entanglements are  zero.  Vice-versa, if $P_{11}=P_{12}=1/2$, then we have the singlet state $\Psi= 1/\sqrt{2}(|01\rangle -|10\rangle)$ which is  the most entangled one.

We see from Fig.\ref{Fig:2qubit} that the relation between the probability of the state transfer and the entanglement in the two-node system is very transparent. Moreover, we will see that relations between  probabilities of  state transfers and  entanglements in more complicated systems are very similar. However, the system with $N>2$ requires additional analysis in order  to find such geometrical configuration of the spin system which allows the HPSTs among all nodes (or, may be, among some of them).   Algorithms allowing us to perform this analysis for the rectangular system of four nodes and for the eight-node system with spins placed in the corners of the parallelepiped  are represented in  Appendices \ref{App:4nodes} and \ref{App:8nodes}.

%%%%%%%%%%%%%%
\section{HPSTs and entanglements  in the system of four nodes with rectangular geometry}
\label{Section:HPST}
\label{Section:four_nodes}
\begin{figure*}[!htb]
\noindent    
\resizebox{50mm}{!}{\includegraphics[width=5cm,
angle=0]{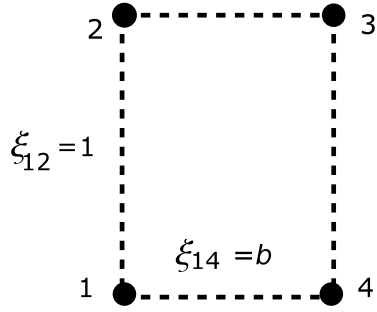}
%{rectangle.eps}
}
  \caption{The rectangular system of four nodes. Here $\xi_{12}=1$  while $\xi_{14}$ is a  parameter: $\xi_{14}=b$. We will use parameter $\delta=1/b^3$ instead of $b$.
  }
  \label{Fig:rectangle}
\end{figure*}

We consider the rectangle shown in Fig.\ref{Fig:rectangle}. 
Here $\xi_{12}=1$ while
 the length of another
side is a parameter of the system, $\xi_{14}=b$.   
However, it is more convenient for us to introduce parameter 
\begin{eqnarray}\label{delta}
\delta=\frac{1}{ b^3}, 
\end{eqnarray}
 instead of $b$. Due to the symmetry of the spin system, 
matrix $D$ in  eq.(\ref{H1})) reads:
\begin{eqnarray}
&&
D=\left(
\begin{array}{cccc}
2\tilde \Gamma&1&d_{13}&d_{14}\cr
1&2\tilde \Gamma&d_{14} & d_{13}\cr
d_{13}&d_{14}&2\tilde \Gamma&1\cr
d_{14} &d_{13}&1&2\tilde \Gamma
\end{array}
\right),\;\;\;
\tilde \Gamma=1+d_{13} + d_{14}.
\end{eqnarray}
The structure of the matrix $D$ helps us to guess  the following set of independent
normalized  eigenvectors:
\begin{eqnarray}
&&
u_1=1/2 (1\;\;-1\;\;1\;\;-1)^T,\;\;u_2=1/2 (1\;\;1\;\;1\;\;1)^T,\\\nonumber
&&
u_3=1/2 (1\;\;-1\;\;-1\;\;1)^T,\;\;u_4=1/2 (1\;\;1\;\;-1\;\;-1)^T
\end{eqnarray}
 with the appropriate eigenvalues:
\begin{eqnarray}
&&
\lambda_1=2\tilde \Gamma-1-d_{14}+d_{13},\;\;\;
\lambda_2=2\tilde \Gamma+1+d_{14}+d_{13},\\\nonumber
&&
\lambda_3=2\tilde \Gamma-1+d_{14}-d_{13},\;\;\;
\lambda_4=2\tilde \Gamma+1-d_{14}-d_{13}.
\end{eqnarray}
It is remarkable that 
\begin{enumerate}
\item
these eigenvectors do not depend on $b$ and
\item
any component of any eigenvector is either 1 or -1.
\end{enumerate}
Namely  the latter property guaranties the HPSTs among all nodes of the four-node system.

Due  to the symmetry of the rectangular cluster, it is enough to consider the case with initial excited state in the first node, i.e. $k_0=1$. 
Eq.(\ref{P^N}) yields explicit expressions for the probabilities $P_{1j}$:
\begin{eqnarray}\label{b0}
P_{11}&=&\frac{1}{4}\Big(1+\cos \tau\Big(\cos(d_{14} \tau) + \cos(d_{13} \tau)\Big) + \cos(d_{14} \tau) \cos(d_{13} \tau)\Big),\\\nonumber
P_{12}&=&\frac{1}{4}\Big(1-\cos \tau\Big(\cos(d_{14}\tau) + \cos(d_{13}\tau)\Big) + \cos(d_{14}\tau) \cos(d_{13} \tau)\Big),\\\nonumber
P_{13}&=&\frac{1}{4}\Big(1+\cos \tau\Big(\cos(d_{14} \tau) - \cos(d_{13} \tau)\Big) - \cos(d_{14} \tau) \cos(d_{13} \tau)\Big),\\\nonumber
P^{(4)}_{14}&=&\frac{1}{4}\Big(1+\cos \tau\Big(-\cos(d_{14} \tau) + \cos(d_{13} \tau)\Big) - \cos(d_{14} \tau) \cos(d_{13} \tau)\Big).
\end{eqnarray}
The explicit formulae relating  the coupling constants   $d_{1j}$ and $b$ depend on the direction of the external magnetic field in accordance with the definition of $d_{ij}$, see eq.(\ref{Dij_dl}). We consider two following cases:
\begin{enumerate}
\item
The external magnetic field is perpendicular to the rectangle, so that 
\begin{eqnarray}\label{perp}
&&
d_{14}=\delta,\;\;\;d_{13}=\frac{1}{(1+b^2)^{3/2}}.
\end{eqnarray}
\item
The external magnetic field is directed along the side $b$ of the rectangle,  so that 
\begin{eqnarray}\label{along_b}
&&d_{14}=-2\delta,\;\;\;d_{13}=\left(1-\frac{3 b^2}{1+b^2}\right)(1+b^2)^{-3/2}.
\end{eqnarray}
\end{enumerate}

Remark that both cases  (\ref{perp}) and (\ref{along_b}) allow equality $|d_{14}|=1$ if  $b=b_0=1$ for case (\ref{perp}) and $b=b_0=2^{1/3}$ for  case (\ref{along_b}). Rectangles with these special  values of $b$ may not provide HPSTs to all nodes. In fact, 
the appropriate  expressions for $P_{1j}$ become simpler in this case:
 \begin{eqnarray}\label{b1}
 P_{11}&=&\frac{1}{4}\left(1+\cos^2 \tau + 2 \cos \tau \cos\left(d_{13}\tau\right)\right),\\\nonumber
 P_{12}&=&P_{14}=\frac{\sin^2 \tau}{4}\le \frac{1}{4}<P_0=0.9,\\\nonumber
 P_{13}&=&\frac{1}{4}\left(1+\cos^2 \tau - 2 \cos \tau \cos\left(d_{13}\tau\right)\right).
 \end{eqnarray}
 We see that in this case $\max P_{12}=\max P_{14} =1/4$, i.e. HPST exists only between 1st and 3rd nodes.
%This means that the square configuration of four spins does not  %provide HPST among all of them in the case (\ref{perp}). %However, 
Any rectangular configuration with $b\neq b_0$  provides HPSTs among all nodes, although the  appropriate time interval ${\cal{T}}$ may be long. 
%Similarly, the rectangular geometry in the case (\ref{along_b}) %which does not provide HPSTs among all nodes is related with the %following ration of sides: $r_{12}/r_{14}=2^{1/3}$.

%%%%%%%%%%%
\subsection{Relationships between probabilities and double negativities}
\label{Section:along} 
As we noted above, any rectangle with $b\neq b_0$ provides HPST among all nodes. However, in general, the time interval ${\cal{T}}$ is long. An important problem is to find such values of $b$ which provide relatively short time interval ${\cal{T}}$. This problem is solved in Appendix \ref{App:4nodes} for the rectangular configuration of four nodes,  where we have found the values of $b$ such that ${\cal{T}}\le 10$ and ${\cal{T}}\le 15$ for the case with magnetic field perpendicular to the plane of rectangle:
\begin{eqnarray}\label{perp_delta}
\delta\in [5.56, 9.62]\;\;\;{\mbox{for }} {\cal{T}}\le 10,\\\nonumber
\delta\in [5.56, 17.79]\;\;\;{\mbox{for }} {\cal{T}}\le 15.
\end{eqnarray}
 and values of $b$ such that  ${\cal{T}}\le 3.5$ and ${\cal{T}}\le 6$ for the case with magnetic field directed along the side $b$:
  \begin{eqnarray}\label{along_delta}
&&
\delta\in [2.62,6.08] \;\;\;{\mbox{for }} {\cal{T}}\le 3.5,\\\nonumber
&&
\delta\in [2.32,6.08]\cup [14.89,30.63]\;\;\;{\mbox{for }} {\cal{T}}\le 6.
\end{eqnarray} 
Thus we see that ${\cal{T}}$ is longer in the first case (magnetic field perpendicular to the plane of the rectangle), i.e. the second case is more preferable for the organization of the HPSTs.

\begin{figure*}[!htb]
\noindent    
\resizebox{160mm}{!}{\includegraphics[width=10cm,angle=270]{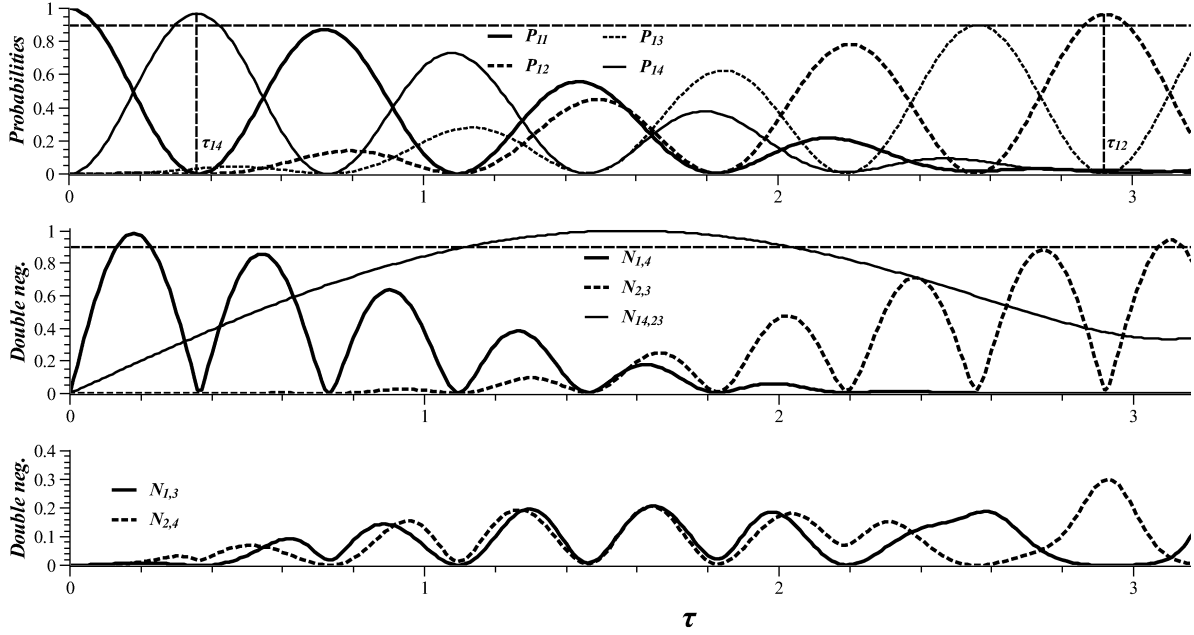}
%{al_P.eps}
}
  %\end{center}
  \caption{Four-node system with the  external field directed along  $\xi_{14}$. ($a$) The  probabilities  $P_{1i}$, $(\tau_{14},\bar P_{14})=(0.36,0.97)$, $(\tau_{12},\bar P_{12})=(2.92,0.96)$, $(\tau_{13},\bar P_{13})=(3.29,0.96)$;  ($b$) the double negativities which provide HPSTs in the system; ($c$) the two-node double negativities which are not associated with HPSTs; 
  }
  \label{Fig:al_P_C}
\end{figure*}

To demonstrate  the qualitative relationship  between the probabilities of the state transfers and the double negativities (see eqs.(\ref{PPT1},\ref{PPT2}-\ref{PPT4})) we show their graphs corresponding to the case (\ref{along_b}) and $\delta=4.3$  in Fig.\ref{Fig:al_P_C} during the interval $T=[0, 3.5]$. We see that the whole interval may be separated into three parts. During the first part, $0\le \tau \lesssim 1.1$, the probabilities $P_{11}$ and $P_{14}$ have big amplitudes. The  associated big amplitude double negativity   is $N_{1,4}$, showing that the first and the 4th nodes are most entangled during this interval. During the last part, $2.2 \lesssim \tau \le 3.5$, the
probabilities   $P_{12}$ and $P_{13}$ have big amplitudes. The appropriate big amplitude double negativity  is $N_{2,3}$ showing that the second and the third nodes are most entangled during this  interval.  The middle part, $1.1\lesssim \tau \le 2.2$,  is characterized by the small values of $P_{1j}$ and considerable double negativity  $N_{14,23}$ showing us that two opposite sides (namely sides  1-4 and 2-3) of the rectangle are well entangled. 
Two more double negativities, $N_{1,2}$ and $N_{3,4}$, are also considerable during the middle interval but they are not represented in the figure because their role is  equivalent to the role of $N_{14,23}$.
The double negativities  $N_{1,3}$ and $N_{2,4}$ remain small during the whole  interval $T$, see Fig.\ref{Fig:al_P_C}$(c)$. We conclude that the time interval  ${\cal{T}}=\tau_{13}\approx 3.29$.

%%%%%%%%%%%%%%%
\section{Three-dimensional spin system of eight nodes  with HPSTs among all of them}
\label{Section:eight_nodes}
\begin{figure*}[!htb]
\noindent    
\resizebox{300mm}{!}{\includegraphics[width=30cm,
angle=0]{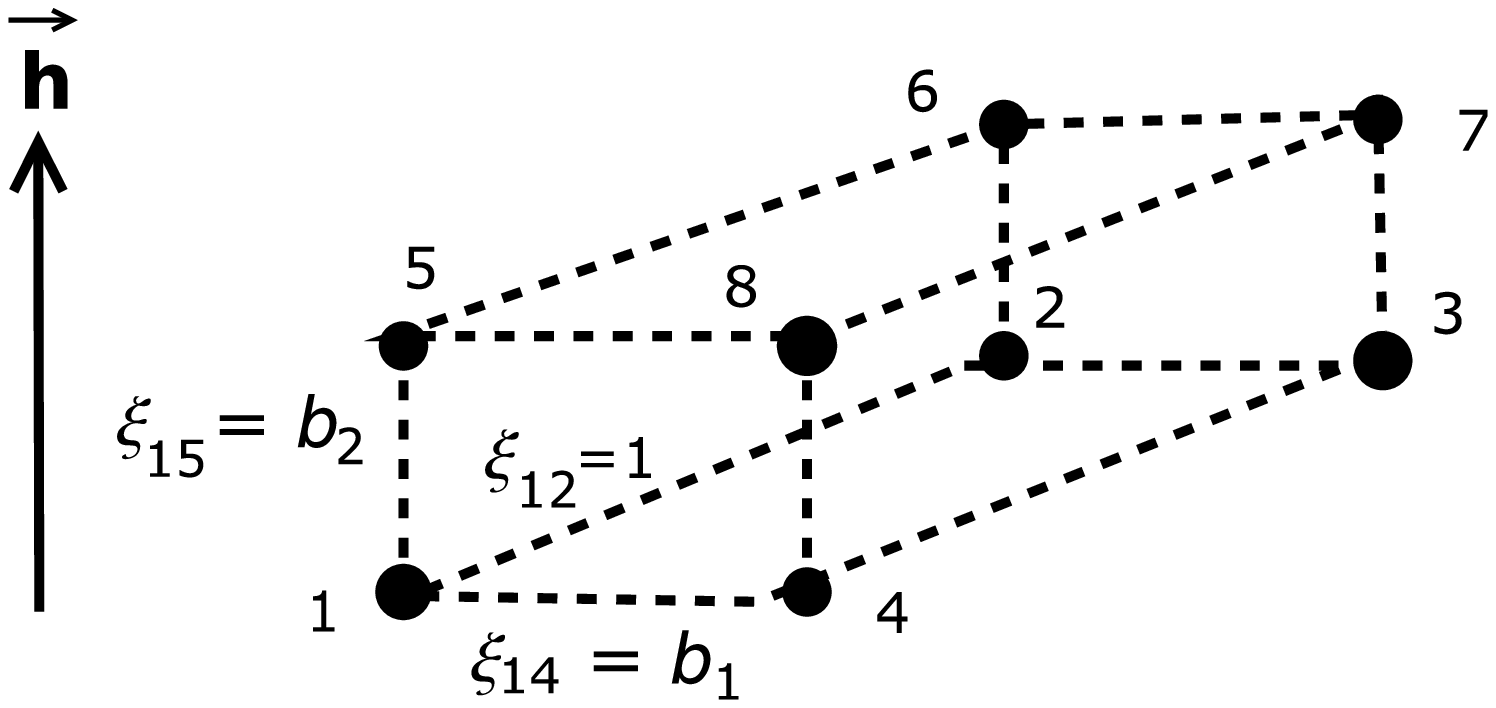}
%{parallelep.eps}
}
  \caption{The three-dimensional eight-node system. The  external magnetic field $\vec h$ is directed along the side  $\xi_{15}$. Here  $\xi_{12}=1$, while two other sides are parameters of this configuration: $\xi_{14}=b_1$, $\xi_{15}=b_2$.  We will use parameters  $\delta_i=1/b_i^3$, $i=1,2$ instead of $b_1$ and $b_2$.
  }
  \label{Fig:parallelep}
\end{figure*}

In this section we consider the parallelepiped with spin-1/2 nodes in its corners, see Fig.\ref{Fig:parallelep}.
Matrix $D$ in eq.(\ref{H1}) has the following block structure:
\begin{eqnarray}
&&
D=\left(
\begin{array}{cccc}
R_1&R_2&R_3&R_4\cr
R_2&R_1&R_4&R_3\cr
R_3&R_4&R_1&R_2\cr
R_4&R_3&R_2&R_1
\end{array}
\right), \\\nonumber
&&
R_1=\left(
\begin{array}{cc}
2\tilde \Gamma&1\cr
1&2\tilde \Gamma
\end{array}
\right),\;\;\;R_2=\left(
\begin{array}{cc}
d_{13}&d_{14}\cr
d_{14}&d_{13}
\end{array}
\right),\;\;\;R_3=\left(
\begin{array}{cc}
d_{15}&d_{16}\cr
d_{16}&d_{15}
\end{array}
\right),\;\;\;R_4=\left(
\begin{array}{cc}
d_{17}&d_{18}\cr
d_{18}&d_{17}
\end{array}
\right),\\\nonumber
&&
d_{13}=(1+b_1^2)^{-3/2},\;\;\;d_{14}=b_1^{-3},\;\;\;
d_{15}=-2 b_2^{-3},\\\nonumber
&& d_{16}=\left(1-3\frac{b_2^2}{1+b_2^2}\right)(1+b_2^2)^{-3/2},\;\;\;
d_{17}=\left(1-3\frac{b_2^2}{1+b_2^2+b_1^2}\right)(1+b_2^2+b_1^2)^{-3/2},\\\nonumber
&&
d_{18}=\left(1-3\frac{b_2^2}{b_2^2+b_1^2}\right)(b_2^2+b_1^2)^{-3/2},\;\;\tilde \Gamma=\sum_{i=2}^8d_{1j},\;\;d_{12}=1.
\end{eqnarray}
Again, the structure of the matrix $D$ allows  us to find the following set of normalized independent eigenvectors:
\begin{eqnarray}\label{8_eigenvectors}
&&
u_1=\frac{1}{2\sqrt{2}}(1 \;1\;1\;1\;1\;1\;1\;1)^T,\;\;\;
u_2=\frac{1}{2\sqrt{2}}(1 \;1\;1\;1\;-1\;-1\;-1\;-1)^T,\\\nonumber
&&
u_3=\frac{1}{2\sqrt{2}}(1 \;1\;-1\;-1\;1\;1\;-1\;-1)^T,\;\;\;
u_4=\frac{1}{2\sqrt{2}}(1 \;1\;-1\;-1\;-1\;-1\;1\;1)^T,\\\nonumber
&&
u_5=\frac{1}{2\sqrt{2}}(1 \;-1\;1\;-1\;1\;-1\;1\;-1)^T,\;\;\;
u_6=\frac{1}{2\sqrt{2}}(1 \;-1\;1\;-1\;-1\;1\;-1\;1)^T,\\\nonumber
&&
u_7=\frac{1}{2\sqrt{2}}(1 \;-1\;-1\;1\;1\;-1\;-1\;1)^T,\;\;\;
u_8=\frac{1}{2\sqrt{2}}(1 \;-1\;-1\;1\;-1\;1\;1\;-1)^T.
\end{eqnarray}
 Similar to the case of four nodes, the eigenvectors do not depend on $b_i$, $i=1,2$.
 The correspondent set of  eigenvalues reads:
\begin{eqnarray}\label{8_eigenvalues}
&&
\lambda_1=3\tilde \Gamma,\;\;\;
\lambda_2=3\tilde \Gamma-2(d_{15}+d_{16}+d_{17}+d_{18}),\\\nonumber
&&
\lambda_3=3\tilde \Gamma-2(d_{13}+d_{14}+d_{17}+d_{18}),\;\;\;
\lambda_4=3\tilde \Gamma-2(d_{13}+d_{14}+d_{15}+d_{16}),\\\nonumber
&&
\lambda_5=3\tilde \Gamma-2(1+d_{14}+d_{16}+d_{18}),\;\;\;
\lambda_6=3\tilde \Gamma-2(1+d_{14}+d_{15}+d_{17}),\\\nonumber
&&
\lambda_7=3\tilde \Gamma-2(1+d_{13}+d_{16}+d_{17}),\;\;\;
\lambda_8=3\tilde \Gamma-2(1+d_{13}+d_{15}+d_{18}),\;\;\;
\end{eqnarray}

Due to the symmetry of our cluster, it is enough to study the case with the initial excited state in the first node, i.e. $k_0=1$ similar to Sec.\ref{Section:four_nodes}. 
The expressions (\ref{P^N}) for the probabilities $P_{1j}$ in terms of the coupling constants $d_{ij}$  are rather complicated so that we do not represent them here. Note, however, that the cube does not allow the HPSTs among  all nodes. In fact, in the case $b_1=b_2=1$ one has
\begin{eqnarray}
P_{12}&=&P_{14}=\frac{\sin^2(2 \tau)}{16}\le \frac{1}{16},\\\nonumber
P_{11}&=&\frac{1}{32}\left(
7+\cos(4\tau)+8\cos \frac{\tau}{4\sqrt{2}} \left(\cos \tau + \cos(2 \tau) \cos\frac{\tau}{4\sqrt{2}}\right)+ \right.\\\nonumber
&&\left.
4\left(\cos \tau + \cos(3 \tau)\right) \cos\frac{3 \tau}{4\sqrt{2}}
\right),\\\nonumber
P_{13}&=&\frac{1}{32}\left(
7+\cos(4\tau)-8\cos \frac{\tau}{4\sqrt{2}} \left(\cos \tau - \cos(2 \tau) \cos\frac{\tau}{4\sqrt{2}}\right)-\right.\\\nonumber
&&\left.
4 \left(\cos \tau + \cos(3 \tau)\right) \cos\frac{3 \tau}{4\sqrt{2}}
\right),\\\nonumber
P_{15}&=&\frac{\sin^2 \tau}{8}\left(
3+\cos(2\tau) +4\cos \tau \cos\frac{3\tau}{4\sqrt{2}}
\right),\\\nonumber
P_{16}&=&P_{18}=\frac{1}{32}\left(
3+\cos(4 \tau) -2\cos\frac{(\sqrt{2}-8)\tau}{4}-
2\cos\frac{(\sqrt{2}+8)\tau}{4}
\right)\le \frac{1}{4},\\\nonumber
P_{17}&=&\frac{\sin^2 \tau}{8}\left(
3+\cos(2\tau) -4\cos \tau \cos\frac{3\tau}{4\sqrt{2}}
\right),
\end{eqnarray}
so that the probabilities  $P_{12}$, $P_{14}$, $P_{16}$ and $P_{18}$ may not approach $P_0=0.9$.

The study of the HPSTs and entanglements in this system requires  finding such subspace of the two-dimensional space of positive parameters  $\delta_1$ and $\delta_2$ that any pair $(\delta_1,\delta_2)$ from this subspace  provides  the HPSTs among all nodes. This  computational problem will not be considered in this paper in the full extend. Instead, we consider an example.  Namely, we will show (see  Appendix 
 \ref{App:8nodes} for details) that the HPSTs among all nodes are possible for $\delta_1=9$ and $\delta_2=26.20$ during the $\tau$-interval  ${\cal{T}}\le 25$.

\begin{figure*}[!htb]
\noindent    
\resizebox{130mm}{!}{\includegraphics[width=10cm,angle=270]{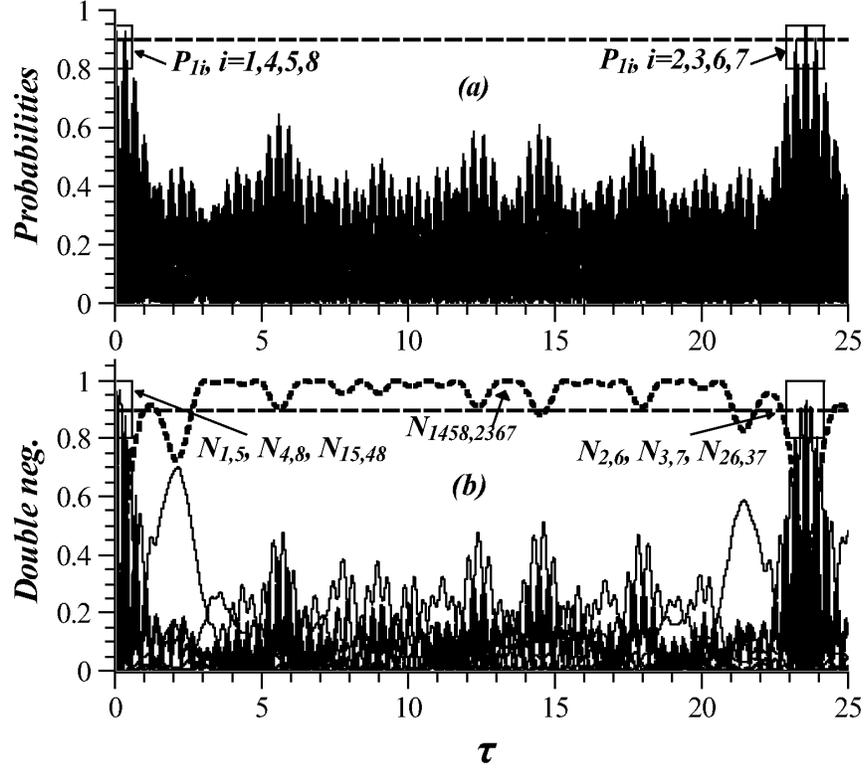}
%{3D_P1.eps}
}
 % \resizebox{70mm}{!}{\includegraphics[width=5cm,angle=270]
 %  {minMax_acr_40.eps}}
  \caption{ Eight-node system. The probabilities and double negativities corresponding to $\delta_1=9$, $\delta_2\equiv \delta=26.20$. The HPSTs take place during the first and  the last parts of the time interval $0\le \tau \le 25$.
 }
  \label{Fig:3D_P1}
\end{figure*}

\begin{figure*}[!htb]
\noindent    
\resizebox{130mm}{!}{\includegraphics[width=10cm,angle=270]{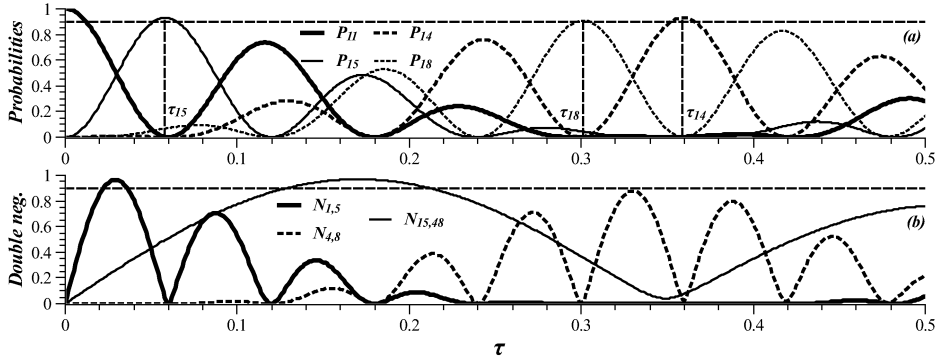}
%{3D_P2.eps}
}
 % \resizebox{70mm}{!}{\includegraphics[width=5cm,angle=270]
 %  {minMax_acr_40.eps}}
  \caption{Eight-node system. The probabilities and double negativities corresponding to $\delta_1=9$, $\delta_2\equiv \delta=26.20$ and HPSTs during the time interval $0\le \tau \le 0.5$; $(\tau_{15},\bar P_{15})=(0.06,0.93)$,
   $(\tau_{18},\bar P_{18})=(0.30,0.91)$,  $(\tau_{14},\bar P_{14})=(0.36,0.93)$ 
 }
  \label{Fig:3D_P2}
\end{figure*}

\begin{figure*}[!htb]
\noindent    
 \resizebox{130mm}{!}{\includegraphics[width=10cm,angle=270]{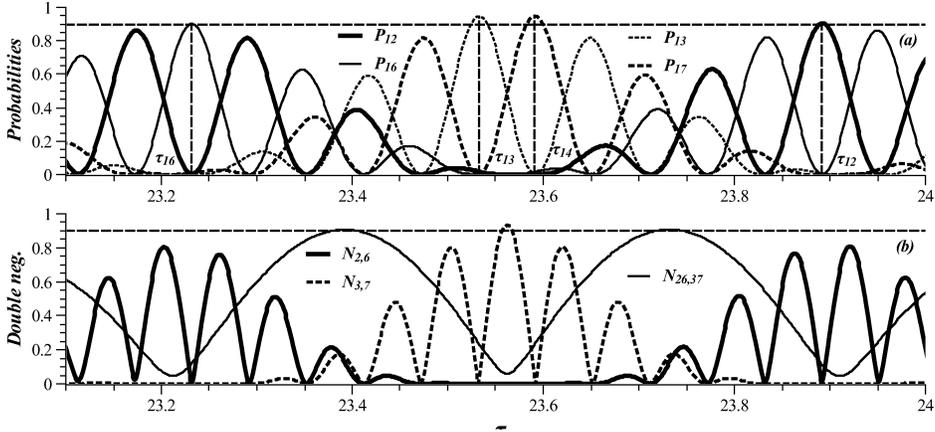}
%{3D_P3.eps}
}
 % \resizebox{70mm}{!}{\includegraphics[width=5cm,angle=270]
 %  {minMax_acr_40.eps}}
  \caption{Eight-node system. The probabilities and double negativities corresponding to $\delta_1=9$, $\delta_2\equiv \delta=26.20$ and HPSTs during the time interval $23.1\le \tau \le 24$; $(\tau_{16},\bar P_{16})=(23.23,0.90)$,
   $(\tau_{13},\bar P_{13})=(23.53,0.95)$,  $(\tau_{17},\bar P_{17})=(23.59,0.95)$ ,  $(\tau_{12},\bar P_{12})=(23.89,0.91)$ 
 }
  \label{Fig:3D_P3}
\end{figure*}

We demonstrate that the relationship between probabilities and double negativities is very similar to one considered in Sec.\ref{Section:four_nodes}. 
 For this purpose, let us refer to  Figs.\ref{Fig:3D_P1}-\ref{Fig:3D_P3}. 
 Figs.\ref{Fig:3D_P1}$(a)$ and $(b)$ collect all probabilities 
 and all double negativities involved into the state transfer 
 process. Similar to the four-node system considered in Sec.\ref{Section:four_nodes}, we select three parts of the whole interval $T=[0,25]$: $0<\tau \lesssim 0.5$, $0.5\lesssim \tau \lesssim 23$ and  $23 \lesssim \tau \lesssim 24$. It is clear from  Fig.\ref{Fig:3D_P1}$(a)$, that the HPSTs 
 take place in the first  and in the third  parts of $T$. All probabilities of the state transfers are not high 
 during the second part of the above interval. Amplitudes of the  probabilities $P_{1i}$, $i=1,4,5,8$, are big during the first  part of the interval, while amplitudes of the  probabilities $P_{1i}$, $i=2,3,6,7$, are big  during the third part of the interval $T$. Similarly, 
 Fig.\ref{Fig:3D_P1}$(b)$ shows that  double negativities $N_{1,5}$, $N_{4,8}$ and $N_{15,48}$  are 
 significant during  the first part of the interval $T$ while double negativities $N_{2,6}$, $N_{3,7}$ and $N_{26,37}$  are significant during the third part of this interval.  One more double negativity   $N_{1458,2367}$ is significant 
 during the second part  of the interval $T$ and is not high during the first and the last parts. This means that namely  $N_{1458,2367}$ is 
 responsible for the HPSTs from the plane 1-4-5-8 to the plane 2-3-6-7. 
 
The probabilities and double negativities during the first and the third  parts of the interval $T$ are 
represented  in Figs.\ref{Fig:3D_P2} and 
\ref{Fig:3D_P3}   respectively in more details. We show only 
those probabilities whose  amplitudes 
exceed  the value $P_0=0.9$. One can see from 
Fig.\ref{Fig:3D_P2}$(a,b)$ that the HPSTs to the $4$th, $5$th and $8$th 
nodes occur during the interval $0\le \tau \le \tau_{14}=0.36$.  Functions 
$N_{1,5}$ and $N_{4,8}$ provide the  HPSTs between the first and the $5$th and
between the 4th and the 8th  nodes respectively, while $N_{15,48}$ 
provides the HPST from the side 1-5  to the 
side  4-8. 

Similarly, we see from Fig.\ref{Fig:3D_P3}$(a,b)$ that the HPSTs to 
the $2$nd, $3$rd, $6$th and $7$th nodes occur during the time interval 
$(\tau_{16}=23.23)\le \tau \le (\tau_{12}=23.89)$.  Functions $N_{2,6}$ and $N_{3,7}$ provide the
HPSTs between the $2$nd and the $6$th  and between the 3rd and the 7th  nodes 
respectively, while $N_{26,37}$ provides HPST from the side 2-6  to the side  3-7. 
We also conclude that ${\cal{T}}=\tau_{12}=23.89$.

%%%%%%%%%%%
\section{Conclusions}

We have derived simple relations between
the probabilities of the excited state transfers to different 
nodes of the spin system and the entanglements between different 
parts of this system described by PPT criterion  for the 
arbitrary spin system in the external magnetic field with single 
excited spin and any Hamiltonian preserving the number of excitations, such as $XXZ$ Hamiltonian. Although similar relations 
for the concurrence (which is  a measure of the entanglement 
between two  nodes in accordance to Wootters criterion)  has been
found \cite{AOPFP,GMT}, 
the PPT criterion allows one to involve entanglements between two
arbitrary  subsystems of nodes  which is  important for the 
systems with HPSTs among many nodes as has been illustrated in 
this paper. 

We have found examples of 4- and 8-node spin systems which provide HPSTs among all nodes. 
We have seen that the HPSTs between two subsystems 
require the high  entanglement between them. This is  illustrated  in
 all examples considered in this paper:    two-node, four-node and eight-node nuclear spin systems, see Figs.\ref{Fig:2qubit},
\ref{Fig:al_P_C}, \ref{Fig:3D_P1}-\ref{Fig:3D_P3} respectively.

In all spin systems considered in this paper, the HPSTs among many nodes  are possible due to the remarkable property of the eigenvectors of the matrix $D$ (see eq.(\ref{H1})): all their elements are real,  equal by absolute value and deffer only by sign.  It is important that four- and eight-nodes systems with this property have simple geometrical configuration: either rectangle (four nodes) or  parallelepiped (eight nodes). It is possible to construct the higher dimensional eigenvectors with this property using the following simple algorithm.

 Consider the eigenvector spaces with the above mentioned  property:
all elements of all eigenvectors are real and have the same absolute value. Suppose that we have $M$-dimensional space ${\cal{B}}_M$ of such vectors. Then we may construct $2M$ dimensional space 
by the formula
\begin{eqnarray}
{\cal{B}}_{2M} = 1/\sqrt{2} \;\; \Big((1,1)\otimes {\cal{B}}_M \cup {(1,-1)}\otimes {\cal{B}}_M\Big).
\end{eqnarray}
Thus we are able to construct $2^s$-dimensional basis with the above property. However it is difficult to  find appropriate spin configurations   different from those which have been described above. Of course, it is quite possible that the HPSTs among many nodes may be arranged using completely different mechanism allowing one, for instance,  to handle the coupling constants in complicated spin systems \cite{BCMS,TPVH}, which is one of the open problems for further study. 

This work is supported by Russian Foundation for Basic Research through the grant 07-07-00048 and by the Program of the Presidium of RAS No.18.

%%%%%%%%%%%%%%%%
\appendix
 %%%%%%%%%%%%
\section{Wootters criterion. Derivation of eq.(\ref{CP})}
\label{App:Wootters}
Our calculations are based on  eqs.(\ref{C_ij}-\ref{rho_red_ij}).
By construction, the reduced density matrix  $\rho^{red}_{ij}$ defined by eq.(\ref{rho_red_ij})  takes a simple $4\times 4$-dimensional   form  in the following  basis
\begin{eqnarray}
|10\rangle, \;\;|01\rangle, \;\;|00\rangle, \;\;|11\rangle,
\end{eqnarray}
where the first and the second  elements are assotiated with the $i$th and the $j$th nodes respectively.
One has
\begin{eqnarray}
\rho^{red}_{ij}=\left(\begin{array}{cccc}
a_{ii} &a_{ij}&0&0\cr
a_{ji} &a_{jj}
&0&0\cr
0&0&\sigma_{ij}&0\cr
0&0&0&0
\end{array}\right),\;\;\;\sigma_{ij}=\sum_{{n=1}\atop{n\neq i,j}}^N
a_{nn}= \sum_{{n=1}\atop{n\neq i,j}}^N 
P_{k_0n}=1-P_{k_0i}-P_{k_0j}.
\end{eqnarray}
Then
\begin{eqnarray}
\tilde\rho^{red}_{ij}=\left(\begin{array}{cccc}
a_{jj} &a_{ji}&0&0\cr
a_{ij}&a_{ii}
&0&0\cr
0&0&0&0\cr
0&0&0&\sigma_{ij}
\end{array}\right).
\end{eqnarray} 
Direct calculation shows that the matrix $\hat\rho_{ij}$ (\ref{hat_rho}) has only one non-zero eigenvalue: $\lambda_1=4 |a_{ij}|^2 = 4 P_{k_0i} P_{k_0j} $, which yields eq.(\ref{CP}).

%Then
%\begin{eqnarray}\label{CP}
%C_{ij}=2 |a_{ij}|=2\sqrt{P_{k_0i} P_{k_0j}},\;\;i\neq j. 
%\end{eqnarray}
%This relation is valid for the system with any number of spins %and for any Hamiltonian commuting with $I_z$ \cite{AOPFP,GMT}.

 %%%%%%%%%%%%
\section{PPT criterion. Derivation of eq.(\ref{PPT1})}
\label{App:PPT}

 We derive formulae (\ref{PPT1}) in this section. First of all, in order to calculate   $N_{i_1\dots i_{M_1},j_1\dots j_{M_2}}$
 one needs the reduced density matrix 
$\rho_{i_1\dots i_{M_1},j_1\dots j_{M_2}}$ calculated in accordance with eq.(\ref{DN_rho_red}) where
\begin{eqnarray}
A=\{i_1,\dots,i_{M_1}\},
\;\;B=\{j_1,\dots,j_{M_2}\}.
\end{eqnarray}
This is $2^{M_1+M_2}\times 2^{M_1+M_2}$ matrix. However, most of elements of this matrix are zeros. All nonzero elements are collected in the $K\times K$ ($K=M_1+M_2+M_1M_2+1$)  block on the diagonal of $\rho_{i_1\dots i_{M_1},j_1\dots j_{M_2}}$. This block corresponds to the subspace spanned by the following set of  vectors (we use notations equivalent to ones introduced for basis (\ref{basis_ordered})):  
\begin{eqnarray}
|i_n\rangle,\;\;
|j_m\rangle,\;\;|0\rangle,\;\;
 |i_n j_m\rangle,\;\;n=1,\dots, M_1,\;\;m=1,\dots, M_2.
\end{eqnarray}
We refer to this block as $\tilde \rho_{i_1\dots i_{M_1}j_1\dots j_{M_2}}$:
\begin{eqnarray}
\tilde \rho_{i_1\dots i_{M_1}j_1\dots j_{M_2}} =
\left(
\begin{array}{ccc}
Q_{i_1\dots i_{M_1}j_1\dots j_{M_2}}& 0_I&0_{II}\cr
0_{I}^T&\sigma_{i_1\dots i_{M_1}j_1\dots j_{M_2}}& 0_{III}\cr
0_{II}^T&0_{III}^T&0_{IV}
\end{array}
\right),
\end{eqnarray}
where $Q_{i_1\dots i_{M_1}j_1\dots j_{M_2}}$ is $(M_1+M_2)\times (M_1+M_2)$ square matrix
\begin{eqnarray}
&&
Q_{i_1\dots i_{M_1}j_1\dots j_{M_2}}=
\left(
\begin{array}{cc}
R_{i_1\dots i_{M_1}}&R_{i_1\dots i_{M_1}j_1\dots j_{M_2}}\cr
 R^T_{i_1\dots i_{M_1}j_1\dots j_{M_2}}& R_{j_1\dots j_{M_2}}
\end{array}
\right)
,\\\nonumber
&&R_{i_1\dots i_{M_1}}=\left(
\begin{array}{ccc}
a_{i_1i_1}&\cdots& a_{i_1i_{M_1}}\cr
\cdots&\cdots&\cdots\cr
a_{i_{M_1}i_1}&\cdots& a_{i_{M_1}i_{M_1}}
\end{array}
\right),\;\;
R_{j_1\dots j_{M_2}}=\left(
\begin{array}{ccc}
a_{j_1j_1}&\cdots& a_{j_1j_{M_2}}\cr
\cdots&\cdots&\cdots\cr
a_{j_{M_2}j_1}&\cdots& a_{j_{M_2}j_{M_2}}
\end{array}
\right),\\\nonumber
&&
R_{i_1\dots i_{M_1}j_1\dots j_{M_2}}=\left(
\begin{array}{ccc}
a_{i_1j_1}&\cdots& a_{i_1j_{M_2}}\cr
\cdots&\cdots&\cdots\cr
a_{i_{M_1}j_1}&\cdots& a_{i_{M_1}j_{M_2}}
\end{array}
\right)\equiv \left(
\begin{array}{c}
q_{i_1;j_1\dots j_{M_2}}\cr
\cdots \cr
q_{i_{M_1};j_1\dots j_{M_2}}\cr
\end{array}
\right),\\\nonumber
&&
q_{i_k;j_1\dots j_{M_2}}=(a_{i_kj_1}\;\;cdots \;\;a_{i_kj_{M_2}}),
\end{eqnarray}
$0_I$ is the column of $M_1+M_2$ zeros, $0_{II}$ is the $ (M_1+M_2)\times M_1M_2$ zero matrix, $0_{III}$ is the row of $M_1M_2$ zeros,
$0_{IV}$ is the $M_1M_2\times M_1M_2$ square matrix  of zeros.

Now we have to transpose the matrix $\rho_{i_1\dots i_{M_1}j_1\dots j_{M_2}}$ with respect to the nodes $i_1,\dots,i_{M_1}$. The nonzero diagonal block of the resulting matrix is associated with  transposition of the block $\tilde\rho_{i_1\dots i_{M_1}j_1\dots j_{M_2}}$ with respect to the nodes $i_1,\dots,i_{M_1}$:
\begin{eqnarray}
\tilde\rho^{i_1\dots i_{M_1}}_{i_1\dots i_{M_1}j_1\dots j_{M_2}} =
\left(
\begin{array}{ccc}
\tilde Q_{i_1\dots i_{M_1}j_1\dots j_{M_2}}& 0_I&0_{II}\cr
0_{I}^T&\sigma_{i_1\dots i_{M_1}j_1\dots j_{M_2}}& R_{I}\cr
0_{II}^T&R_I^+&0_{IV}
\end{array}
\right),
\end{eqnarray}
where $^+$ means hermitian conjugation,
\begin{eqnarray}
&&
\tilde Q_{i_1\dots i_{M_1}j_1\dots j_{M_2}}=
\left(
\begin{array}{cc}
R^T_{i_1\dots i_{M_1}}&0_{V}\cr
 0_V^T& R_{j_1\dots j_{M_2}}
\end{array}
\right), \;\;R_I=(q_{i_1;j_1\dots j_{M_2}}\dots q_{i_{M_1};j_1\dots j_{M_2}}),
\end{eqnarray}
$0_{V}$ is the  $M_1\times M_2$ zero matrix.
By construction, both blocks  $R^T_{i_1\dots i_{M_1}}$ and $R_{j_1\dots j_{M_2}}$ are density matrices and, as a consequence, have non-negative eigenvalues. 
Thus, all eigenvalues of the matrix $\tilde Q^{red}_{i_1\dots i_{M_1}j_1\dots j_{M_2}}$ are non-negative. For this reason, looking for the negative eigenvalues of the matrix 
$\tilde \rho^{i_1\dots i_{M_1}}_{i_1\dots i_{M_1}j_1\dots j_{M_2}}$, 
 we stay with the  matrix 
\begin{eqnarray}
\left(
\begin{array}{cc}
\sigma_{i_1\dots i_{M_1}j_1\dots j_{M_2}}& R_{I}\cr
R_I^+&0_{IV}
\end{array}
\right).
\end{eqnarray}
Its characteristics equation reads:
\begin{eqnarray}
(-\lambda)^{M_1M_2-1}\left(\lambda^2 -\lambda \sigma_{i_1\dots i_{M_1}j_1\dots j_{M_2}}-\sum_{n=1}^{M_1}\sum_{m=1}^{M_2}
|a_{i_nj_m}|^2\right)=0,
\end{eqnarray} 
which has two nonzero roots one of which is negative:
\begin{eqnarray}
\lambda_1=\frac{1}{2}\left(\sigma_{i_1\dots i_{M_1}j_1\dots j_{M_2}}-
\sqrt{\sigma_{i_1\dots i_{M_1}j_1\dots j_{M_2}}^2 + 4 \sum_{n=1}^{M_1}\sum_{m=1}^{M_2} |a_{i_n j_m}|^2}\right).
\end{eqnarray}
Consequently,
\begin{eqnarray}
N_{i_1\dots i_{M_1},j_1\dots j_{M_2}}=2|\lambda_1|,
\end{eqnarray}
which generates the formulae (\ref{PPT1}).

%%%%%%%%%%%%%%%
\section{Optimization of the rectangular system. Values of the parameter $\delta$ providing the HPSTs among all nodes}
\label{App:4nodes}
 There are four functions  characterizing HPSTs in the system of four spin-1/2 nodes:
\begin{eqnarray}\label{P}
P_{1i}(\tau),\;\;\;i=1,2,3,4.
\end{eqnarray}
As for the entanglements, the situation is  more complicated. We have the following lists of all possible double negativities  in the four-node system:
\begin{eqnarray}\label{Ent11}
&&
{\cal{N}}_{1,1}\equiv 
%{\mbox{entanglement between two nodes}}=
 \{N_{i,j},\;\;i,j=1,2,3,4,\;\;i\neq j\},\\\label{Ent22}
&&
{\cal{N}}_{2,2}\equiv
%{\mbox{entanglement between different pairs of nodes}}=
\{N_{12,34},\;\;N_{13,24}, \;\;N_{14,23}\},\\\label{Ent13}
&&
{\cal{N}}_{1,3}\equiv
%{\mbox{entanglement between given node
% and the rest of the system}}= \\\nonumber
%&&\hspace{6cm}
\{N_{1,234},\;\;N_{2,134}, \;\;N_{3,124},\;\;N_{4,123}\}.
\end{eqnarray}
%It was shown in \cite{VW}
%that there is an hierarchy of double negativities
% so that entanglements from the list ${\cal{C}}_{1,1}$ are 
%minimal. 
We will show that ${\cal{N}}_{1,1}$ and ${\cal{N}}_{2,2}$ are responsible for  the HPSTs among all nodes.

Our system has one geometrical parameter $b$ or $\delta$ (see eq.(\ref{delta})) completely describing the rectangular  geometry.
We represent an  algorithm allowing one to find such values of the parameter $\delta$  which provide inequality (\ref{HPST})  with  the short  time interval ${\cal{T}}$ (\ref{tT}) .  For this purpose we consider the functions $P_{1i}$ and $N_{i,j}$ as functions of two arguments, $\tau$ and $\delta$. Let us fix some time interval $T$ and check whether the state may be transferred with high probability from the 1st node to any other node and whether the entanglements between any two nodes are significant  during this time interval.  For this purpose we construct two following  functions:
\begin{eqnarray}
 \label{F^P}
&&
F^{P}(\delta,T)=\lim_{\Delta \tau\to 0} F^{P}(\delta,T,\Delta \tau),
\\\label{F^N}
&&
F^{N}(\delta,T)=\lim_{\Delta \tau\to 0} F^{N}(\delta,T,\Delta \tau),
 \end{eqnarray}
where
\begin{eqnarray}\label{Delta_t_4}\label{Delta_t}
 &&
 F^{P}(\delta,T,\Delta \tau)=\min\limits_{k=1,2,3,4}\Big[\max\limits_{i=0,\dots,K}
 P_{1k}(\tau_i,\delta)\Big],\\\nonumber
 &&
 F^{N}(\delta,T,\Delta \tau)=\min\limits_{{n,m=1,2,3,4}\atop{n\neq m}}\Big[\max\limits_{i=0,\dots,K}
 N_{n,m}(\tau_i,\delta)\Big],\;\;\tau_i=i \Delta\tau,\;\;\Delta\tau=\frac{T}{K},\;\;K\in\NN.
\end{eqnarray}
 %\begin{eqnarray}\label{F^P} && %F^{P}(\delta,T)=\min\limits_{k=1,2,3,4}\Big[\max\limits_{\tau\in %[0,T]} 
% P_{1k}(\tau,\delta)\Big],\\\label{F^N} 
% && 
% F^{N}(\delta,T)=
% \min\limits_{{i,j=1,2,3,4}\atop{i\neq j}}
% \Big[\max\limits_{\tau\in [0,T]}
% N_{i,j}(\tau,\delta)\Big].
%\end{eqnarray}
The
 HPSTs among all nodes of the system are possible if there is such $\delta=\delta_0$ that 
\begin{eqnarray}
F^{P}(\delta_0,T)\ge P_0.
\end{eqnarray}
Then 
\begin{eqnarray}
{\cal{T}}\le T.
\end{eqnarray}
Function $F^N$ tells us how significant is the entanglement between any two nodes of the system. 
%Hereafter we take $P_0=0.9$.

\begin{figure*}[!htb]
\noindent    
\resizebox{70mm}{!}{\includegraphics[width=5cm,angle=270]{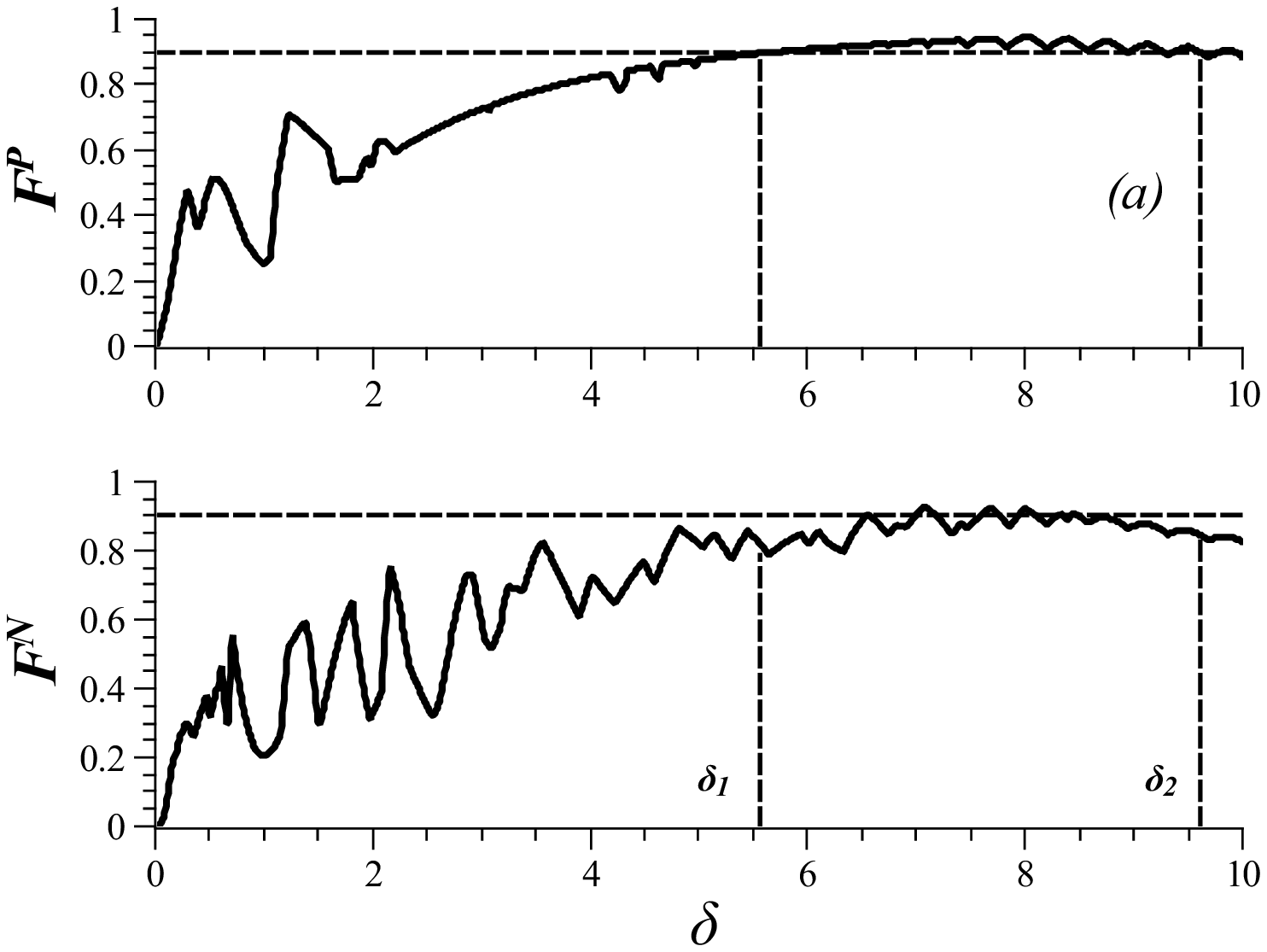}
%{minMax_acr_10.eps}
}
    \hfill 
\resizebox{70mm}{!}{\includegraphics[width=5cm,angle=270]{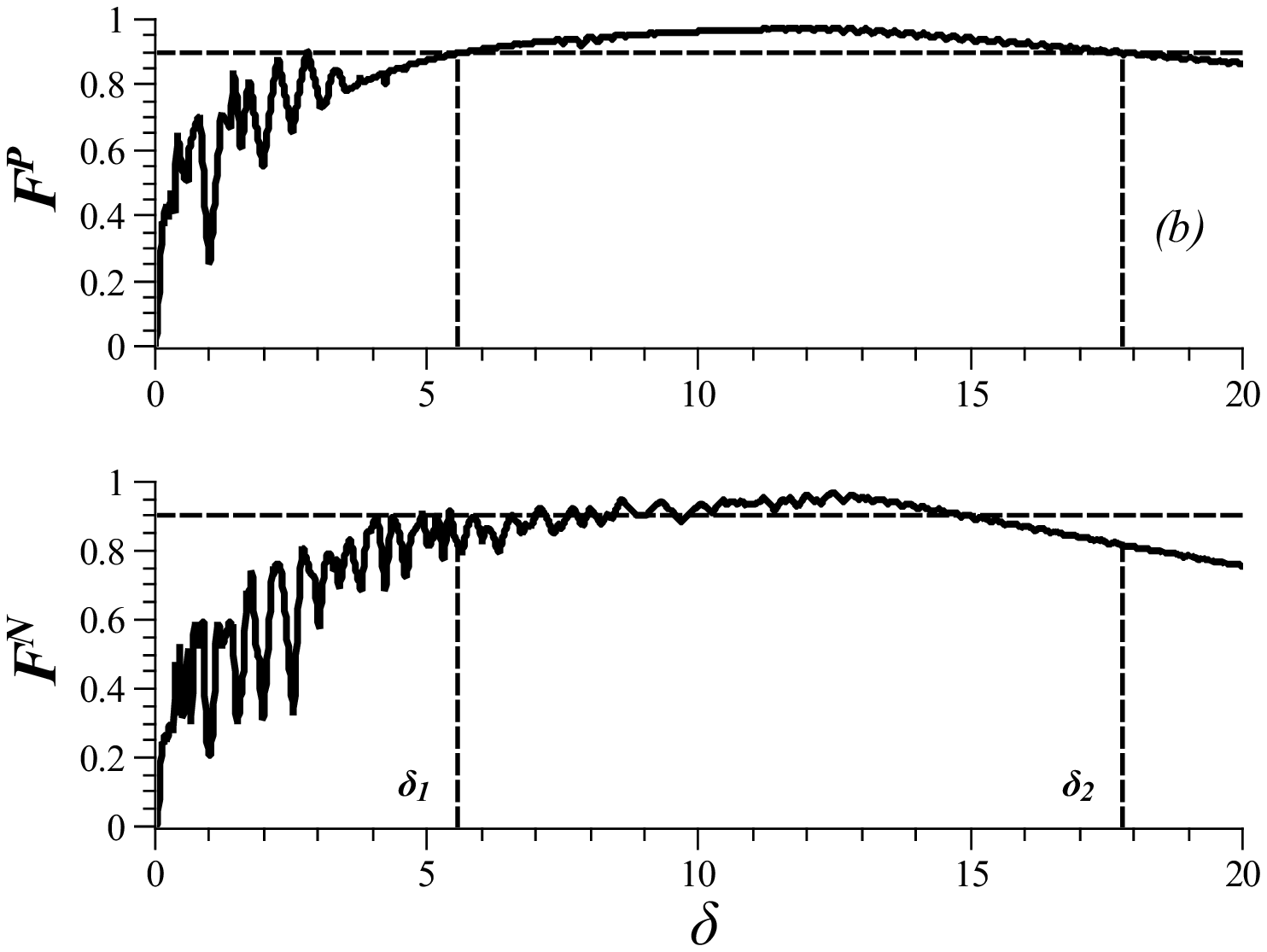}
%{minMax_acr_15.eps}
}
   \hfill  
 % \resizebox{70mm}{!}{\includegraphics[width=5cm,angle=270]
 %  {minMax_acr_40.eps}}
  \caption{Four-node system with the external field  perpendicular to the plane of  the rectangle. Functions $F^P(\delta,T,0.01)$ and $F^N(\delta,T,0.01)$.  $(a)$ $T=10$, $\delta_1=5.56$, $\delta_2=9.62$; $(b)$ $T=15$, $\delta_1=5.56$, $\delta_2=17.79$.
  }
  \label{Fig:min_max}
\end{figure*}

First, we consider the case when the  external magnetic field is perpendicular to the rectangle.
Functions $F^P(\delta,T,\Delta \tau)$ and $F^N(\delta,T,\Delta \tau)$ with $\Delta\tau=0.01$ corresponding to the  intervals  $T=10$ and $15$ are represented in  Fig.\ref{Fig:min_max}.
The HPSTs among all nodes are possible for the parameter $\delta$ inside of the intervals (\ref{perp_delta}).
%\begin{eqnarray}
%\delta\in [5.56, 9.62]\;\;\;{\mbox{for }} T=10,\\\nonumber
%\delta\in [5.56, 17.79]\;\;\;{\mbox{for }} T=15.
%\end{eqnarray} 

The interval of $\delta$ corresponding to the HPSTs among all nodes   increases with increase in $T$. If $T$ is big enough (we have found that $T \gtrsim  40$), then the HPSTs among all nodes exist even for  $\delta<1$. If $\delta=\delta_0=1$, then $F^P$ may not exceed $1/4$ in accordance with eqs.(\ref{b1}).
Fig.\ref{Fig:min_max} shows  that the function $F^N\sim 0.8 \div 0.9$  when $F^P\gtrsim 0.9$. This means, that  the HPSTs among all nodes during the interval $T$ entangle  any two   nodes in this case, i.e. functions  $N_{i,j}$ (for $i,j=1,2,3,4$, $i\neq j$)  must have big amplitudes  during the time interval $T$.

\begin{figure*}[!htb]
\noindent    
\resizebox{70mm}{!}{\includegraphics[width=5cm,angle=270]{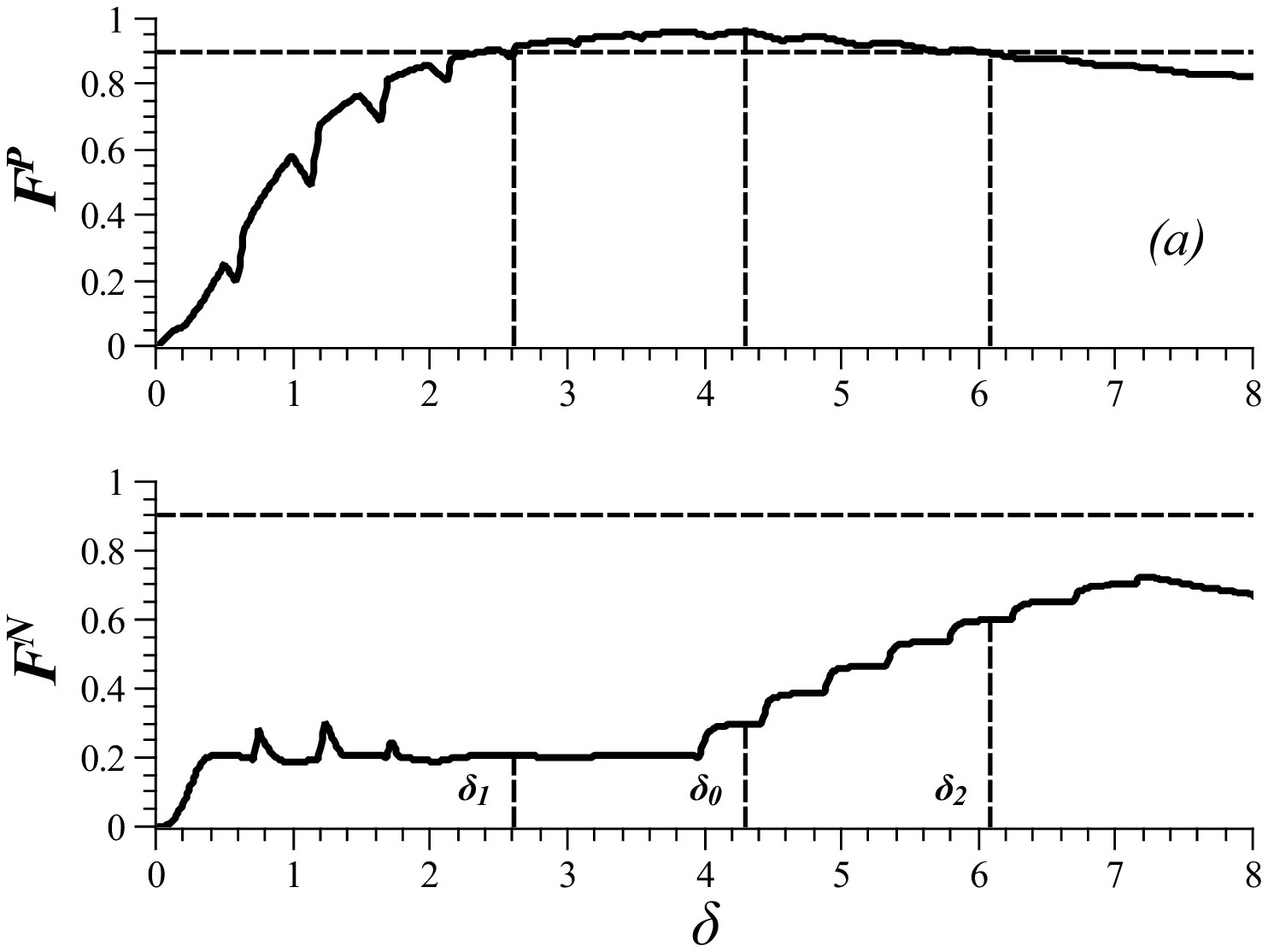}
%{minMax_al_3_5.eps}
}
    \hfill 
\resizebox{70mm}{!}{\includegraphics[width=5cm,angle=270]{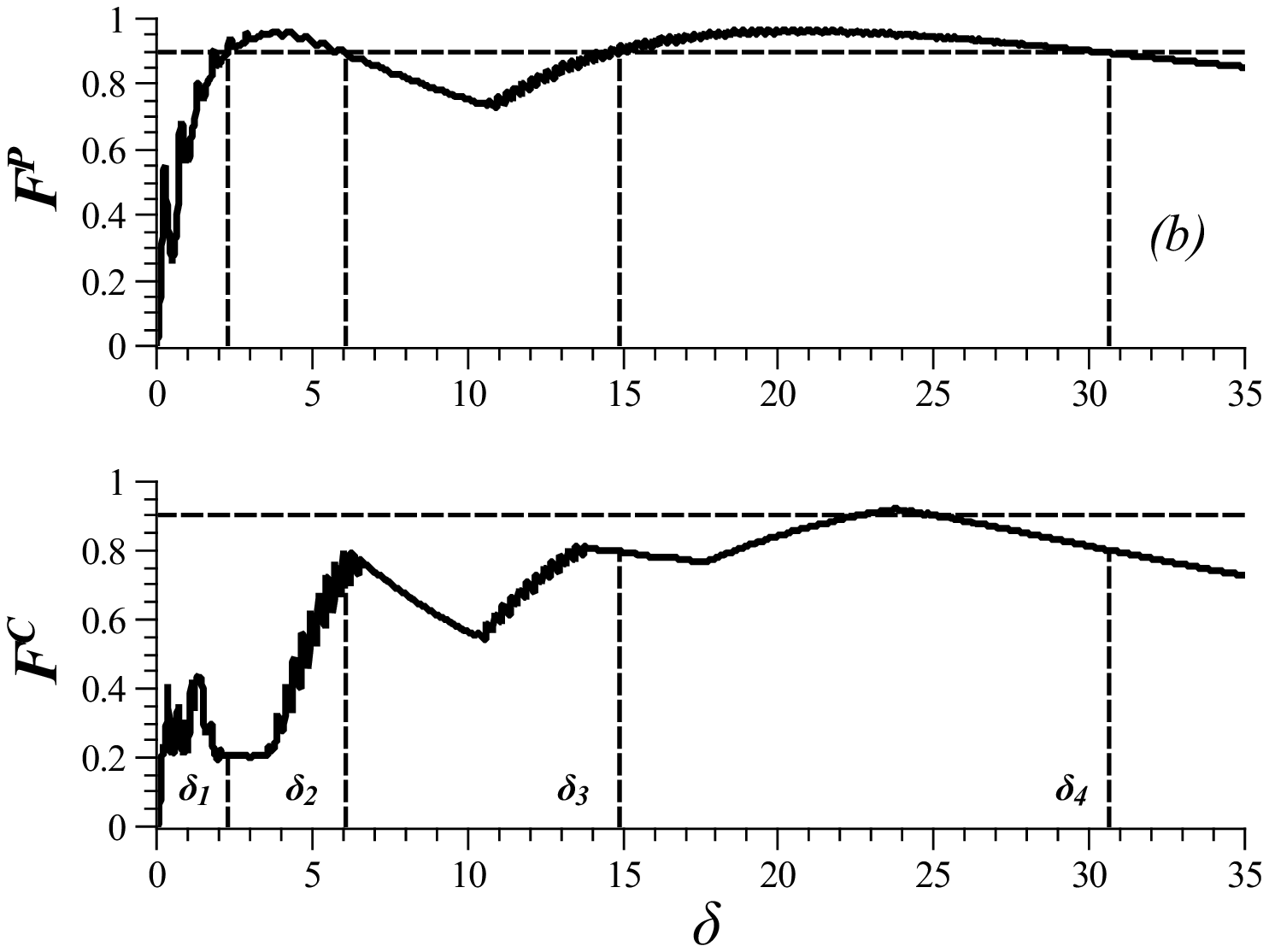}
%{minMax_al_6.eps}
}
  \caption{Four-node system with the external field  directed  along $\xi_{14}$. Functions $F^P(\delta,T)$ and $F^N(\delta,T)$; $(a)$ $T=3.5$, $\delta_1=2.62$, $\delta_2=6.08$, $\delta_0=4.3$. $F^P(\delta_0,3.5)=0.96$ $(b)$ $T=6$, $\delta_1=2.32$, $\delta_2=6.08$, $\delta_3=14.89$, $\delta_4=30.63$.
  }
  \label{Fig:min_max_b}
\end{figure*}

Similarly, the functions $F^{P}(\delta,T,\Delta \tau)$ and $F^{N}(\delta,T,\Delta \tau)$ for the case with the magnetic field directed along $\xi_{14}$ are represented in  Fig.\ref{Fig:min_max_b} for, $T=3.5$, $\Delta \tau=0.01$ and $T=6$, $\Delta\tau=0.001$. 
The HPSTs among all nodes are  possible for $\delta$ inside of the intervals (\ref{along_delta}). 
%\begin{eqnarray}
%&&
%\delta\in [2.62,6.08] \;\;\;{\mbox{for }} T=3.5,\\\nonumber
%&&
%\delta\in [2.32,6.08]\cup [14.89,30.36]\;\;\;{\mbox{for }} T=6.
%\end{eqnarray} 
In this case the  function $F^N$ does not necessary take a big value   when $F^P\gtrsim 0.9$. This means, that not any two nodes  must be entangled in order to provide the HPSTs among all nodes  during the time interval $T$.  
Note that
exceptional value of $\delta$ is $\delta=\delta_0=1/2$, when  $F^P$ does not exceed $1/4$ in accordance with eqs.(\ref{b1}).

Comparison of Figs.\ref{Fig:min_max}$(a)$ and \ref{Fig:min_max_b}$(a)$ shows that the second case (i.e. external field is directed along $\xi_{14}$) is more preferable for the organization of the HPSTs  because appropriate interval $T$ is almost three times  shorter.

Figs.\ref{Fig:min_max} and \ref{Fig:min_max_b} show that the intervals of the parameter $\delta$ providing the HPSTs among all nodes increase with increase in $T$, i.e. the system becomes  more "stable" with respect to  variations in $b$ (compare  the interval $\delta_1\le \delta\le \delta_2$ in Fig.\ref{Fig:min_max} and the intervals $\delta_1\le \delta\le \delta_2$, $\delta_3\le \delta\le \delta_4$ in Fig.\ref{Fig:min_max_b}).

%To demonstrate the relationship between probabilities and double %negativities (eqs.(\ref{PPT1}-\ref{PPT4}))
% we  consider the system with the external  magnetic field along %$\xi_{14}$, $T=3,5$ and   $\delta=4.3$, %Fig.\ref{Fig:min_max_b}$(a)$.

Finally, remark that   function  (\ref{Delta_t}) are decreasing functions of $\Delta \tau$. Thus, one may expect that the graphs of the functions $F^{P}(\delta,T)$ and $F^{N}(\delta,T)$ are above of the appropriate  curves shown in Figs.\ref{Fig:min_max} and \ref{Fig:min_max_b}. However, further decrease in $\Delta \tau$ negligibly effects the shapes of the above curves in all examples of this section.

%%%%%%%%%%%%%%%%%%
\section{On the optimization of the eight-node system. Values of the parameters $\delta_1$ and $\delta_2$ providing the HPSTs among all nodes}
\label{App:8nodes}
We introduced functions $F^P$ and $F^N$  in Appendix \ref{App:4nodes}, see  eqs.(\ref{F^P},\ref{F^N}). However, we have seen that $F^N$ is not as helpful as $F^P$ in defining the optimal parameter $\delta$. In fact, Figs.\ref{Fig:min_max} and \ref{Fig:min_max_b} show us that $F^N$ may be either big  or small when $F^P\gtrsim 0.9$ reflecting the fact that not all entanglements between two nodes are important for the HPSTs during the interval $T$.
For this reason we consider only $F^P$ in this section:

\begin{eqnarray}\label{F^P_8}
 \label{tt}
&&
F^{P}(\delta,T)=\lim_{\Delta \tau\to 0} F^{P}(\delta,T,\Delta \tau),
 \end{eqnarray}
where
\begin{eqnarray}\label{Delta_t_3D}
 F^{P}(\delta,T,\Delta \tau)=\min\limits_{k=1,\dots,8}\Big[\max\limits_{i=0,\dots,K}
 P_{1k}(\tau_i,\delta)\Big],\;\;\tau_i=i \Delta\tau,\;\;\Delta\tau=\frac{T}{K},\;\;K\in\NN,
\end{eqnarray}
 and $P_{1k}$ are given by  eqs.(\ref{P^N},\ref{8_eigenvectors},\ref{8_eigenvalues}).

\begin{figure*}[!htb]
\noindent    
\resizebox{70mm}{!}{\includegraphics[width=5cm,angle=270]{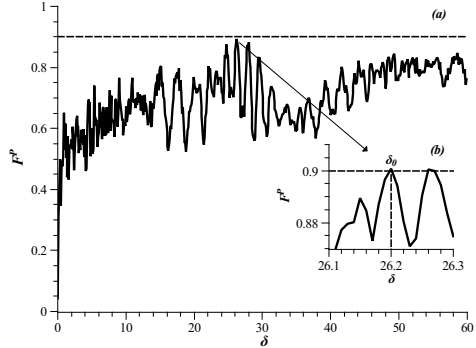}
%{minmax3Ddel9.eps}
} 
  \caption{ Function $F^P(\delta,T,0.01)$ for  $T=25$; $\delta_0=26.20$.
 }
  \label{Fig:minmax3Ddel9}
\end{figure*}

  Function $F^{P}(\delta,T,\Delta \tau)$ 
  is shown in Fig.\ref{Fig:minmax3Ddel9} for $T=25$ and $\Delta \tau=0.01$. 
 We see from  Fig.\ref{Fig:minmax3Ddel9}$(b)$ that the HPSTs among all nodes  are possible, for instance, for $\delta=\delta_0=26.20$.

\end{document}